*Feature Paper*

# Agreement Technologies for Coordination in Smart Cities

**Holger Billhardt [1], Alberto Fernández[1], Marin Lujak[2] and Sascha Ossowski[1,*]**

[1] CETINIA, Univ. Rey Juan Carlos, Spain; {holger.billhardt, alberto.fernandez, sascha.ossowski@urjc.es}
[2] IMT Lille Douai, France; marin.lujak@imt-lille-douai.fr
**\*** Correspondence: sascha.ossowski@urjc.es; Tel.: +34 916.647.485

**Abstract:** Many challenging problems in today's society can be tackled by distributed open systems. This is particularly true for domains that are commonly perceived under the umbrella of Smart Cities, such as intelligent transportation, smart energy grids, or participative governance. When designing computer applications for these domains, it is necessary to account for the fact that the elements of such systems, often called software agents, are usually made by different designers and act on behalf of particular stakeholders. Furthermore, it is unknown at design time when such agents will enter or leave the system, and what interests new agents will represent. To instil coordination in such systems is particularly demanding, as usually only part of them can be directly controlled at runtime. Agreement Technologies refer to a sandbox of tools and mechanisms for the development of such open multiagent systems, which are based on the notion of agreement. In this paper, we argue that Agreement Technologies are a suitable means for achieving coordination in Smart Cities domains, and back our claim through examples of several real-world applications.

## 1. Introduction

The transactions and interactions among people in modern societies are increasingly mediated by computers. From email, over social networks, to virtual worlds, the way people work and enjoy their free time is changing dramatically. The resulting networks are usually large in scale, involving huge numbers of interactions, and are open for the interacting entities to join or leave at will. People are often supported by software components of different complexity to which some of the corresponding tasks can be delegated. In practice, such systems cannot be built and managed based on rigid, centralised client-server architectures, but call for more flexible and decentralised means of interaction.

The field of Agreement Technologies (AT) [1] envisions next-generation open distributed systems, where interactions between software components are based on the concept of agreement, and which enact two key mechanisms: a means to specify the "space" of agreements that the agents can possibly reach, and an interaction model by means of which agreements can be effectively reached. Autonomy, interaction, mobility and openness are key characteristics that are tackled from a theoretical and practical perspective.

Coordination in Distributed Systems is often seen as governing the interaction among distributed processes, with the aim of "gluing together" their behaviour, so that the resulting ensemble shows desired characteristics or functionalities [2]. This notion has also been applied to Distributed Systems made up of software agents. Initially, the main purpose of such multiagent systems was to efficiently perform problem-solving in a distributed manner: both the agents and their rules of interaction were designed together, often in a top-down manner and applying a divide-and-conquer strategy to solve the problem at hand [3]. However, many recent applications of multiagent systems refer to domains where agents, possibly built by different designers and representing different interests, may join and leave the system at a pace that is unknown at design time. It is apparent that coordination in such open multiagent systems requires a different, extended stance on coordination [3].

Application areas that fall under the umbrella of *Smart Cities* have recently gained momentum [4]. Intelligent transportation systems, smart energy grids, or participative governance are just some examples of domains where an improved efficiency of the use of shared urban resources (both physical and informational) can lead to a better quality of life for the citizens. It thus seems evident that new applications in the context of Smart Cities have the potential for achieving significant socio-economic impact.

We believe that applying AT to the domain of Smart Cities may enable the development of novel applications, both with regard to functionality for stakeholders as well as with respect to the level of sustainability of Smart City services. In particular, in this article we discuss how coordination can be achieved in practical applications of multiagent systems, with different levels of openness, by making use of techniques from the sandbox of AT. Section 2 briefly introduces the fields of AT, coordination models, and Smart Cities, and relates them to each other. Section 3 describes several real-world applications, related to the field of Smart Cities, that illustrate how coordination models can be tailored to each particular case and its degree of openness. Section 4 summarises the lessons learnt from this enterprise.

## 2. Background

In this section we introduce the fields of Agreement Technologies and Coordination models and relate them to each other. We then briefly characterise the field of Smart Cities, and argue that Agreement Technologies are a promising candidate to instil coordination in Smart City aplications.

*2.1. Agreement Technologies*

Agreement Technologies (AT) [1] address next-generation open distributed systems, where interactions between software processes are based on the concept of agreement. AT-based systems are endowed with means to specify the "space" of agreements that can be reached, as well as interaction models for reaching agreement and monitoring agreement execution. In the context of AT, the elements of open distributed systems are usually conceived as software agents. There is still no consensus where to draw the border between programs or objects on the one hand and software agents on the other, but the latter are usually characterised by four key characteristics, namely *Autonomy*, *Social ability*, *Responsiveness* and *Proactiveness* [5]. The interactions of a software agent with its environment (and with other agents) are guided by a reasonably complex program, capable of rather sophisticated activities such as reasoning, learning, or planning. Two main ingredients are essential for such multiagent systems based on AT: firstly, a normative model that defines the "rules of the game" that software agents and their interactions must comply with; and secondly, an interaction model where agreements are first established and then enacted. AT can then be conceived as a sandbox of methods, platforms, and tools to define, specify and verify such systems.

The basic elements of the AT sandbox are related to the challenges outlined by Sierra et al. for the domain of Agreement Computing [6], covering the fields of semantics, norms, organisations, argumentation & negotiation, as well as trust & reputation. Still, when dealing with open distributed systems made up of software agents, more sophisticated and computationally expensive models and mechanisms can be applied [7].

The key elements of the field of AT can be conceived of in a tower structure, where each level provides functionality to the levels above, as depicted in Figure 1.

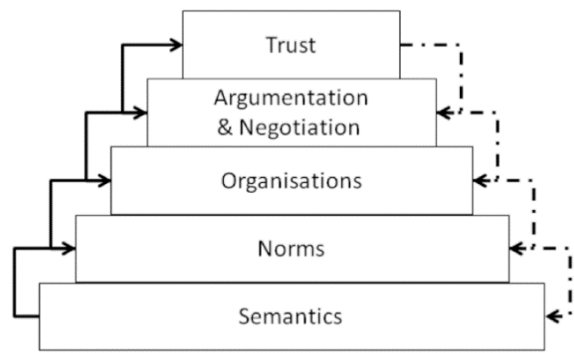

**Figure 1.** AT tower

*Semantic technologies* provide solutions to semantic mismatches through the alignment of ontologies, so agents can reach a common understanding on the elements of agreements. In this manner, a shared multi-faceted "space" of agreements can be conceived, providing essential information to the remaining layers. The next level is concerned with the definition of *norms* determining constraints that the agreements, and the processes leading to them, should satisfy. Thus, norms can be conceived of as a means of "shaping" the space of valid agreements. *Organisations* further restrict the way agreements are reached by imposing organisational structures on the agents. They thus provide a way to efficiently design and evolve the space of valid agreements, possibly based on normative concepts. The *argumentation and negotiation* layer provides methods for reaching agreements that respect the constraints that norms and organisations impose over the agents. This can be seen as choosing certain points in the space of valid agreements. Finally, the *trust and reputation* layer keeps track of as whether the agreements reached, and their executions, respect the constraints put forward by norms and organisations. So, it complements the other techniques that shape the "agreement space", by relying on social mechanisms that interpret the behaviour of agents.

Even though one can clearly see the main flow of information from the bottom towards the top layers, results of upper layers can also produce useful feedback that can be exploited at lower levels. For instance, as mentioned above, norms and trust can be conceived as a priori and a posteriori approaches, respectively, to security [6]. Therefore, in an open and dynamic world it will certainly make sense for the results of trust models to have a certain impact on the evolution of norms. Some techniques and tools are orthogonal to the AT tower structure. The topics of environments [8] and infrastructures [9], for instance, pervade all layers. In much the same way, coordination models and mechanisms are not just relevant to the third layer of Figure 1 , but cross-cut the other parts of the AT tower as well [10]. We will elaborate on this matter in the next subsection.

*2.2. Coordination models*

The notion of coordination is central to many disciplines. Sociologists observe the behaviour of groups of people, identify particular coordination mechanisms, and explain how and why they emerge. Economists are concerned with the structure and dynamics of the market as a particular coordination mechanism; they attempt to build coordination market models to predict its behaviour. Biologists observe societies of simple animals demonstrating coordination without central coordinators; coordination mechanisms inspired from Biology have proven useful to various scientific disciplines. In Organizational Theory, the emphasis is on predicting future behaviour and performance of an organization, assuming the validity of a certain coordination mechanism. From a Computer Science point of view, the challenge is to *design* mechanisms that "glue together" the activities of distributed actors in some efficient manner. However, beyond such high-level conceptions, within the Computer Science field, and even among researchers working on Multiagent Systems, there is no commonly agreed definition for the concept of coordination. An important reason for this are the different *interests* of the designers in coordination mechanisms (micro and/or macro level properties), as well as different level of control that designers have over the elements of the distributed intelligent system (the degree of *openness* of the system), as we will argue in the following [3].

Early work on coordination in MAS focused essentially on (Cooperative) *Distributed Problem Solving*. In this field, it is assumed that a system is constructed (usually from the scratch) out of several intelligent components, and that there is a single designer with *full control* over these agents. In particular, this implies that agents are *benevolent* (as instrumental local goals can be designed into them) and, by consequence, that the designer is capable of imposing whatever interaction patterns are deemed necessary to achieve efficient coordination within the system. Efficiency in this context usually refers to a trade-off between the system's resource consumption and the quality of the solution provided by the system: agents necessarily have only partial, and maybe even inconsistent views of the global state of the problem-solving process, so they need to exchange just enough information to be able to locally take good decisions (i.e., choices that are instrumental with respect to the overall system functionality). Resource consumption is not only measured in terms of computation but also of communication load.

From a *qualitative* perspective, coordination in Distributed Problem-Solving systems can be conceived as a *Distributed Constraint Problem* (see [11] for an example). Agents locally determine individual actions that comply with the constrains (dependencies) that affect them, so as to give rise to "good" global solutions. Alternatively, in *quantitative* approaches the structure of the coordination problem is hidden in the shape of a shared global multi-attribute utility function. An agent has control over only some of the function's attributes, and the global utility may increase/decrease in case there is a positive/negative dependency with an attribute governed by another agent, but these dependencies are hidden in the algorithm that computes the utility function, and are thus not declaratively modelled. Quantitative approaches to coordination can be understood in terms a of a *Distributed Optimization Problem*.

More recent research in the field of MAS has been shifting the focus towards *open* systems, where the assumption of a central designer with full control over the system components no longer holds. This raises interoperability problems that need to be addressed. In addition, the benevolence assumption of Distributed Problem Solving agents needs to be dropped: coordination mechanisms now have to deal with *autonomous*, self-interested behaviour – an aspect that is usually out of the scope of models from the field of Distributed Computing. Approaching agent design in open systems from a *micro-level* perspective means designing an intelligent software entity capable of successful autonomous action in potentially hostile (multiagent) environments. In this context coordination can be defined as "a way of adapting to the environment" [12]: adjusting one's decisions and actions to the presence of other agents, assuming that they show some sort of *rationality*. If the scenario is modelled within a quantitative framework, we are still concerned with multi-attribute utility functions, where only some attributes are controlled by a particular agent, but now there are different utility functions for each agent. The most popular way of characterizing a problem of these characteristics is through (non-constant sum) *Games* [13]. Coordination from a micro-level perspective this boils down to agents applying some sort of "best response" strategy, and potentially leads to some notion of (Nash) equilibrium. From a macro-level perspective, coordination is about designing "rules of the game" such that, assuming that agents act rationally and comply with these rules, some desired properties or functionalities are instilled. In the field of Game Theory, this is termed *Mechanism Design* [13]. In practice, it implies designing potentially complex interaction protocols among the agents, which shape their "legal" action alternatives at each point in time, as well as institutions or infrastructures that make agents abide by the rules [9]. From this perspective instilling coordination in an open multiagent system can be conceived as an act of *governing interaction* within the system.

If the environment is such that agents can credibly commit to mutually binding agreements, coordinating with others comes down to negotiating the terms such commitments. This is where the link to AT becomes evident. *Norms* and *organisations* define and structure the interactions that may take place among agents. The shape of these interactions depends on the particular case, but often they can be conceived as *negotiating* an agreement for a particular outcome of coordination. In addition, depending on the agents' interests, information and structured *arguments* can be provided to make agents converge on such an agreement. Norms and trust can be seen as a priori and a posteriori measures, respectively, that make agents comply with the constraints imposed by norms and organisations.

*2.3. Smart Cities*

There is a broad variety of domains where the potential of AT becomes apparent (see Part VII of [1]). In these domains, the choices and actions of a large number autonomous stakeholders need to be coordinated, and interactions can be regulated, by some sort of intelligent computing infrastructure [9], through some sort of institutions and institutional agents [14], or simply by strategically providing information in an environment with a significant level of uncertainty [15]. The advent of intelligent road infrastructures, with support for vehicle-to-vehicle and vehicle-to-infrastructure communications, make *smart transportation* a challenging field of application for AT, as it allows for a decentralized coordination of individually rational commuters. But also the infrastructure of the electricity grid is evolving, allowing for bidirectional communication among energy producers and consumers. Therefore, in the near future large numbers of households could coordinate and adapt their aggregate energy demand to the supply offered by utilities. AT can also be applied to the domain of *smart energy* in order to integrate large numbers of small-scale producers of renewable energy into the grid infrastructure. In much the same way, *smart governance* can make use of electronic institutions the support citizens, for instance in e process of dispute resolution.

The above are only a few examples of applications and domains that are often referred to under the umbrella of *Smart Cities*. Even though many definitions of that term exist [4], there is still no commonly agreed conception of a smart city. Still, we believe that authors tend to concur that a key challenge of Smart Cities is to improve the efficiency of the use of shared urban resources (both physical and informational) through the use of ICT, so as to improve the quality of life of citizens (see, e.g., [16, 17, 18]). Most of the world's urban areas have a limited space to expand, congestion and contamination seriously affect people's well-being, and a constant and reliable supply of energy is essential for almost all aspects of urban life. Therefore, ICT-based solutions can help to adequately disseminate information and effectively coordinate the urban services and supplies, so as to make urban life more comfortable and efficient.

While initially smart city research had a strong focus on ICT and "smartness", more recently impact indicators of environmental, economic or social *sustainability* have also gained importance [19], so in present days the term *smart sustainable city* is commonly used [20]. This notion underlines that, on the tack to making our cities smarter, preserving the "needs of present and future generations with respect to economic, social and environmental aspects" is of foremost importance [21].

The Internet of Things (IoT) is often considered as crucial in the development of smart cities [22]. It is usually conceived as a global infrastructure, enabling advanced services by interconnecting (physical and virtual) devices based on ICT. Recently, it has been moving from interesting proofs of concept to systemic support for urban processes that generates efficiency *at scale* (see, e.g., [23]). With increasing connectivity between people, data and things based on IoT, the challenge is how to manage and *coordinate* the decisions of a myriad of decision makers in real time considering the scarcity of resources and stochasticity in demand.

We believe that there is a significant potential in applying AT outlined in section 2.1, targeting methods and tools that support the formation and execution of agreements in large-scale open systems, in order to progress towards the vision of smart and sustainable cities mentioned above. In much the same way, it seems straightforward that the efficient discovery, orchestration, and maintenance of services, based largely on data from heterogeneous sensors and all sorts of embedded devices, calls for the application of both scalable and tailorable coordination models. In the following sections, we will focus on different types of assignment problems in the context of sustainable smart cities: we provide examples of how AT-based coordination services mediate the use of scarce resources to the benefit of citizens.

## 3. Applications

In this section we show how the AT paradigm can be applied to achieve coordination in various real-world problems. Depending on the structure and characteristics of each domain, different technologies from the AT sandbox need to be selected and combined so as to meet the requirements for each case. Section 3.1. highlights the use of techniques related to norms and organisations (in particular, auction protocols and market-based control) in an open domain, where flows of autonomous vehicles, controlled by individually rational driver agents, are coordinated through a network of intelligent interactions. Section 3.2 is dedicated to the problem of evacuation guidance in smart buildings where evacuees, suffering from significant levels of uncertainty concerning the state of an emergency, are provided with individualised route recommendations in a coordinated manner. In this context, issues related to situation-awareness and semantics play a major role. Section 3.3. addresses the coordination of fleets of ambulance vehicles. Even though this is a primarily closed scenario, we address it by techniques from the field of AT applying an algorithm that *simulates* multiple concurrent computational auctions. Section 3.4 focuses on coordination of emergency medical services for angioplasty patients – a problem similar to the previous one, even though its internal structure (different types of agents, etc.) leads to a more complex coordination mechanism. Finally, section 3.5. also addresses a coordination problem related to fleet management, but applied to the field of taxi services. Here, we again have a higher degree of openness, as taxis are conceived of as autonomous agents, so coordination needs to be induced by incentives, targeted at influencing the choices of drivers whose actions are not fully determined by organizational rules and protocols.

### *3.1. Coordination of traffic flows through intelligent intersections*

Removing the human driver from the control loop through the use of autonomous vehicles integrated with an intelligent road infrastructure can be considered as the ultimate, long-term goal of the set of systems and technologies grouped under the name of Intelligent Transportation Systems (ITS). Autonomous vehicles are already a reality. For instance, in the *DARPA Grand Challenges*[1] different teams competed to build the best autonomous vehicles, capable of driving in traffic, performing complex manoeuvres such as merging, passing, parking and negotiating with intersections. The results have shown that autonomous vehicles can successfully interact with both manned and unmanned vehicular traffic in an urban environment. In line with this vision, the *IEEE Connected Vehicle initiative*[2] promotes technologies that link road vehicles to each other and to their physical surroundings, i.e., by vehicle-to-infrastructure and vehicle-to-vehicle wireless communications. The advantages of such an integration span from improved road safety to a more efficient operational use of the transportation network. For instance, vehicles can exchange critical safety information with the infrastructure, so as to recognise high-risk situations in advance and therefore to alert drivers. Furthermore, traffic signal systems can communicate signal phase and timing information to vehicles to enhance the use of the transportation network.

In this regard, some authors have recently paid attention to the potential of a tighter integration of autonomous vehicles with the road infrastructure for future urban traffic management. In the *reservation-based* control system [24], an intersection is regulated by a software agent, called *intersection manager* agent, which assigns reservations of space and time to each autonomous vehicle intending to cross the intersection. Each vehicle is operated by another software agent, called *driver* agent. When a vehicle approaches an intersection, the driver requests that the intersection manager reserves the necessary space-time slots to safely cross the intersection. The intersection manager, provided with data such as vehicle ID, vehicle size, arrival time, arrival speed, type of turn, etc., simulates the vehicle's trajectory inside the intersection and informs the driver whether its request is in conflict with the already confirmed reservations. If such a conflict does not exist, the driver stores the reservation details and tries to meet them; otherwise it may try again at a later time. The authors show through simulations that in situations of balanced traffic, if all vehicles are autonomous, their delays at the intersection are drastically reduced compared to traditional traffic lights.

---

[1] *https://en.wikipedia.org/wiki/DARPA_Grand_Challenge*

[2] http://sites.ieee.org/connected-vehicles/ieee-connected-vechicles/ieee-cv-initiative/

In this section we report on our efforts to use different elements of the sandbox of AT to further improve the effectiveness and applicability of Dresner and Stone's approach, assuming a future infrastructure where all vehicles are autonomous and capable of interacting with the regulating traffic infrastructure. We extend the reservation-based model for intersection control at two different levels.

- *Single Intersection*: our objective is to elaborate a new policy for the allocation of reservations to vehicles that takes into account the drivers' different attitudes regarding their travel times.
- *Network of Intersections*: we build a computational market where drivers must acquire the right to pass through the intersections of the urban road network, implementing the intersection managers as competitive suppliers of reservations which selfishly adapt the prices to match the actual demand, and combine the competitive strategy for traffic assignment with the auction-based control policy at the intersection level into an adaptive, market-inspired, mechanism for traffic management of reservation-based intersections.

3.1.1. Mechanism for single intersection

For a single reservation-based intersection, the problem that an intersection manager has to solve comes down to allocating reservations among a set of drivers in such a way that a specific objective is maximised. This objective can be, for instance, minimising the average delay caused by the presence of the regulated intersection. In this case, the simplest policy to adopt is allocating a reservation to the first agent that requests it, as occurs with the first-come first-served (FCFS) policy proposed by Dresner and Stone in their original work. Another work in line with this objective takes inspiration from adversarial queuing theory for the definition of several alternative control policies that aim at minimising the average delay [25]. However, these policies ignore the fact that in the real world, depending on people's interests and the specific situation that they are in, the relevance of travel time may be judged differently: a business person on his or her way to a meeting, for instance, is likely to be more sensitive to delays than a student cruising for leisure. Since processing the incoming requests to grant the associated reservations can be considered as a process of assigning resources to agents that request them, one may be interested in an intersection manager that allocates the disputed resources to the agents that value them the most. In the sequel, we design an auction-based policy for this purpose. In line with approaches from mechanism design, we assume that the more a human driver is willing to pay for the desired set of space-time slots, the more they value the good. Therefore, our policy for the allocation of resources relies on *auctions*.

The first step is to define the resources (or items) to be allocated. In our scenario, the auctioned good is the use of the space inside the intersection at a given time. We model an intersection as a discrete matrix of space slots. Let $S$ be the set of the intersection space slots, and $T$ the set of future time-steps, then the set of items that a bidder can bid for is $I = S \times T$. Therefore, differently from other auction-based approaches for intersection management (e.g. [26]), our model of the problem calls for a *combinatorial auction*, as a bidder is only interested in bundles of items over the set $I$. As Figure 2 illustrates, in the absence of acceleration in the intersection, a reservation request implicitly defines which space slots at which time the driver needs in order to pass through the intersection.

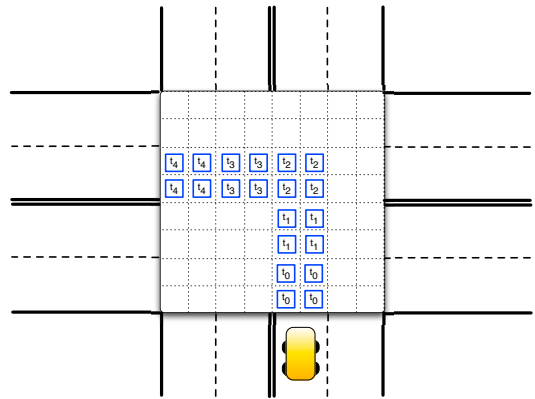

Figure 2. Bundle of items for a reservation request

The bidding rules define the form of a valid bid accepted by the auctioneer. In our scenario, a bid over a bundle of items is implicitly defined by the reservation request. Given the parameters *arrival time*, *arrival speed*, *lane* and *type of turn*, the auctioneer (i.e., the intersection manager) is able to determine which space slots are needed at which time. Thus, the additional parameter that a driver must include in its reservation request is the *value of its bid*, i.e., the amount of money that it is willing to pay for the requested reservation. A bidder is allowed to withdraw its bid and to submit a new one. This may happen, for instance, when a driver that submitted a bid *b*, estimating to be at the intersection at time *t*, realises that, due to changing traffic conditions, it will more likely be at the intersection at time *t'>t*, thus making the submitted bid *b* useless for the driver. The rational thing to do in this case, as the driver would not want to risk being involved in a car accident, is resubmitting the bid with the updated arrival time. However, we require the new bid to be greater than or equal to the value of the previous one. This constraint avoids the situation whereby a bidder "blocks" one or several slots for itself, by acquiring them early and with overpriced bids.

Figure 3 shows the interaction protocol used in our approach. It starts with the auctioneer waiting for bids for a certain amount of time. Once the new bids are collected, they constitute the bid set. Then, the auctioneer executes the algorithm for the winner determination problem (WDP), and the winner set is built, containing the bids whose reservation requests have been accepted. During the WDP algorithm execution, the auctioneer still accepts incoming bids, but they will only be included in the bid set of the next round. The auctioneer sends a *CONFIRMATION* message to all bidders that submitted the bids contained in the winner set, while a *REJECTION* message is sent to the bidders that submitted the remaining bids. Then a new round begins, and the auctioneer collects new incoming bids for a certain amount of time.

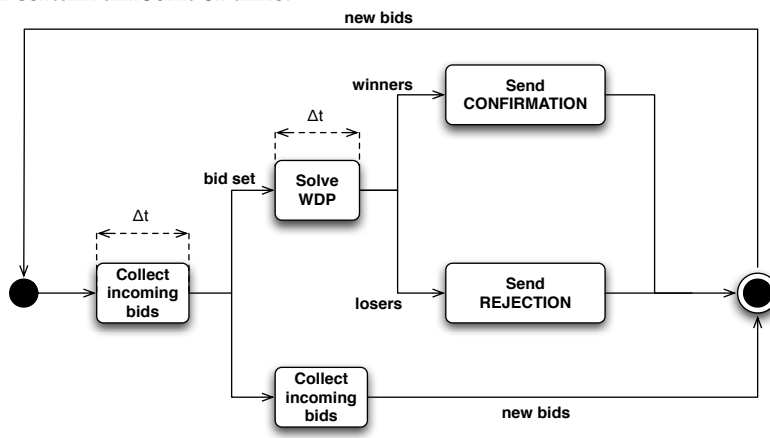

**Figure 3.** Auction policy

Notice that the auction must be performed in real-time, so both the bid collection and the winner determination phase must occur within a specific time window. This implies that optimal and complete algorithms for the WDP are not suitable. Therefore, we use an approximation algorithm with *anytime* properties, i.e. the longer the algorithm keeps executing, the better the solution it finds [27].

We expect our policy based on combinatorial auction (CA) to enforce an inverse relation between the amounts spent by the bidders and their *delay* (the increase in travel time due to the presence of the intersection). That is, the more money a driver is willing to spend for crossing the intersection, the faster will be its transit through it. For this purpose, we designed a custom, microscopic, time- and-space-discrete simulator, with simple rules for acceleration and deceleration [27]. The origin O and destination D of each simulated vehicle are generated randomly. The destination implies the type of turn that the vehicle will perform at the intersection as well as the lane it will use to travel. We create different traffic demands by varying the expected number of vehicles $\lambda$ that, for every O-D pair, are spawned in an interval of 60 seconds, using a Poisson distribution. The bid that a driver is willing to submit is drawn from a normal distribution with mean 100 cents and variance 25 cents, so the agents are not homogeneous in the sense that the amount of money that they are offering differs from one to another.

Figure 4 plots (in logarithmic scale) the relation between travel time and bid value for values of $\lambda$=20, with error bars denoting 9% confidence intervals. It clearly shows a sensible decrease of the delay experienced by the drivers that bid from 100 to 150 cents. The delay reduction tends to settle for drivers that bid more than 1000 cents. Similar results are achieved with lower and higher densities [27]. Notice that even with a theoretically infinite amount of money, a driver cannot experience zero delay when approaching an intersection, as the travel time is influenced by slower potentially "poorer" vehicles in front of it. Extensions to our mechanism that address this problem are subject to future work.

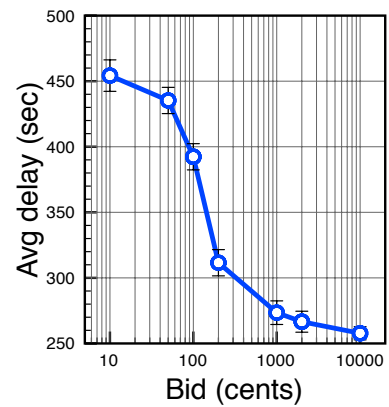

**Figure 4.** Bid-delay relation

We also analysed the impact that such a policy has on the intersection's *average* delay, comparing it to the FCFS strategy. Figure 5 shows that when traffic demand is low, the performance of the CA policy and the FCFS is approximately the same, but as demand grows there is a noticeable increase of the average delay when the intersection manager applies CA. The reason is that the CA policy aims at granting a reservation to the driver that values it the most, rather than maximising the number of granted requests. Thus, a bid *b* whose value is greater than the sum of *n* bids that share some items with *b* is likely to be selected in the winner set. If so, only 1 vehicle will be allowed to transit, while *n* other vehicles will have to slow down and try again. When extending the CA mechanism to multiple intersections, we will try to reduce this "social cost" of giving preference to drivers with a high valuation of time.

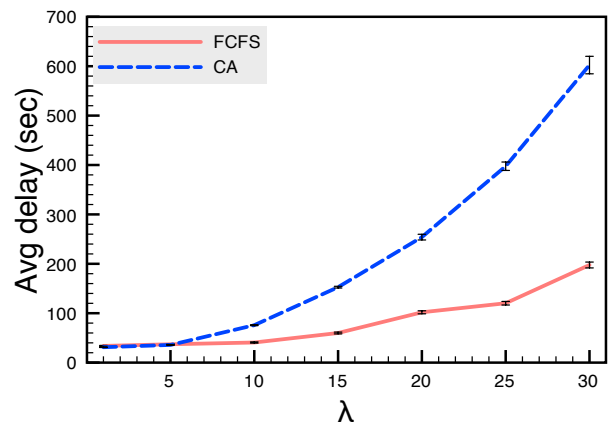

**Figure 5.** Average delay

3.1.2. Mechanisms for multiple intersections

In the previous section we have analysed the performance of an auction-based policy for the allocation of reservations in the *single* intersection scenario. A driver is modelled as a simple agent that selects the preferred value for the bid that will be submitted to the auctioneer. The decision space of a driver in an urban road network with *multiple* intersections is much broader: complex and mutually dependent decisions must be taken such as route choice and departure time selection. Therefore, this scenario opens up new possibilities for intersection managers to affect the behaviour of drivers. For example, an intersection manager may be interested in influencing the collective route choice performed by the drivers, using variable message signs, information broadcast, or individual route guidance systems, so as to evenly distribute the traffic over the network. This problem is called *traffic assignment*. In the following we first evaluate how market-inspired methods can be used within a traffic assignment strategy for networks of reservation-based intersections (CTA strategy). Then, we combine this traffic assignment strategy with the auction-based control policy into an integrated mechanism for traffic management of urban road networks (CTA-CA strategy). Finally, the performance of the different approaches is evaluated.

The complexity of the problem puts limits to coordination approaches based on cooperative multiagent learning [28]. Therefore, our *Competitive Traffic Assignment* strategy (CTA) models each intersection manager as a provider of the resources (in this case, the reservations of the intersection it manages). Each intersection manager is free to establish a price for the reservations it provides. On the other side of the market, each driver is modelled as a buyer of these resources. Provided with the current prices of the reservations, it chooses the route, according to its personal preferences about travel times and monetary costs. Each intersection manager is modelled so as to compete with all others for the supply of the reservations that are traded. Therefore, our goal as market designers is making the intersection managers adapt their prices towards a price vector that accounts for an efficient allocation of the resources.

In CTA, for each incoming $l$, an intersection manager defines the following variables:

- Current price $p^t(l)$: the price applied by the intersection manager to the reservations sold to the drivers that come from the incoming link $l$.
- Total demand $d^t(l)$: the total demand of reservations from the incoming link $l$ that the intersection manager observes at time $t$, given the current price $p^t(l)$, i.e. the number of vehicles that intend to cross the intersection coming from link $l$ at time $t$.
- Supply $s(l)$: the reservations supplied by the intersection manager for the incoming link $l$. It is a constant and represents the number of vehicles that cross the intersection coming from link $l$ that the intersection manager is willing to serve.
- Excess demand $z^t(l)$: the difference between total demand at time $t$ and supply, i.e. $d^t(l) - s(l)$.

We define the price vector $p^t$ as a vector that comprises all prices at time $t$, i.e. the prices applied by all intersection manager to each of its controlled links. In particular, we say that a price vector $p^t$ maps the supply with the demand if the excess demand $z^t(l)$ is 0 for all links $l$ of the network. This price vector, which corresponds to the market equilibrium price, can be computed through a Walrasian auction [29] where each buyer (i.e., driver) communicates to the suppliers (i.e., intersection managers) the route that it is willing to choose, given the current price vector $p^t$. With this information, each intersection manager computes the demand $d^t(l)$ as well as the excess demand $z^t(l)$ for each of its controlled links. Then, each intersection manager adjusts the prices $p^t(l)$ for all the incoming links, lowering them if there is excess supply ($z^t(l) < 0$) and raising them if there is excess demand ($z^t(l) > 0$). The new price vector $p^{t+1}$ is communicated to the drivers that iteratively choose their new desired route on this basis. Once the equilibrium price is computed, the trading transactions take place and each driver buys the required reservations at the intersections that lay on its route.

In order to adapt the Walrasian auction to the traffic domain, we implement a pricing strategy that aims at reaching the equilibrium price but works on a continuous basis, with drivers that leave and join the market dynamically, and with transactions that take place continuously. To reach general equilibrium, each intersection manager applies the following *price update rule*: at time $t$, it independently computes the excess demand $z^t(l)$ and updates the price $p^t(l)$ as follows:

$$p^{t+1}(l) = max\left[\delta, p^t(l) + p^t(l)\frac{z^t(l)}{s(l)}\right]$$

where $\delta$ is the minimum price that the intersection manager charges for the reservations that it sells. As drivers that travel through road network links with low demand shall not incur any costs, for the CTA strategy we choose $\delta=0$.

The *integrated mechanism for traffic management* (CA-CTA) combines the competitive traffic assignment strategy (CTA) with the auction-based policy (CA). Since the intersection manager is the supplier of the reservations that are allocated through the combinatorial auction, it may control the *reserve price* of the auctioned reservations, i.e. the minimum price at which the intersection manager is willing to sell. At time $t$, for each link $l$, CTA-CA simply sets this reserve price to the price $p^t(l)$ computed by the price update rule of the CTA strategy.

The experimental evaluation of the strategies is performed on a hybrid mesoscopic-microscopic simulator, where traffic flow on road segments is modelled at mesoscopic level, while traffic flow inside intersections is modelled at microscopic level. Although our work does not depend on the underlying road network, we chose a topology inspired by the urban road network of the city of Madrid for our empirical evaluation (see Figure 6). The network is characterised by several freeways that connect the city centre with the surroundings and a ring road. Each large dark vertex in Figure 6, if it connects three or more links, is modelled as a reservation-based intersection. In the experiments, we recreate a typical high load situation (i.e., the central, worst part of a morning peak), with more than 11,000 vehicles departing within a time window of 50 minutes (see [27] for details).

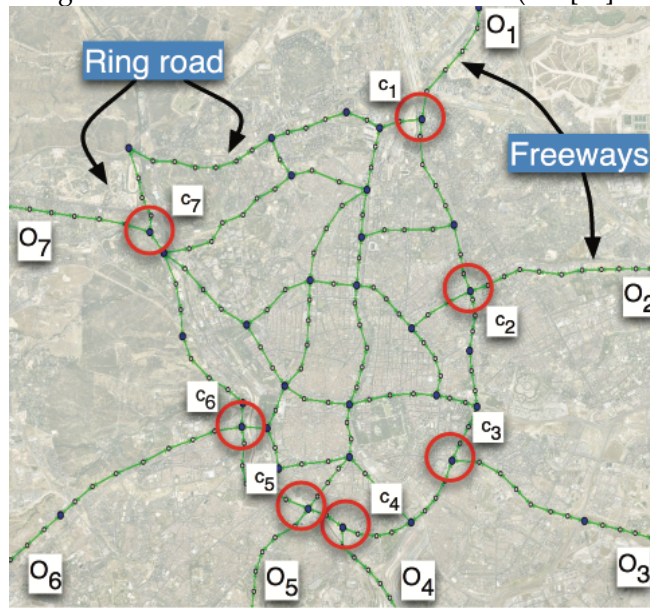

**Figure 6.** Urban road network

We aim at comparing the performance of FCFS, CTA, and CTA-CA. In FCFS, each intersection manager performs combinatorial auctions (without reserve price) in isolation. In this case, the drivers´ route choice model simply selects the route with minimum expected travel time at free flow, since there is no notion of price. For the other strategies, we assume that drivers choose the most preferred route they can afford. Since the prices of links are changing dynamically, a driver continuously evaluates the utility of the route it is following and, in case that a different route becomes more attractive, it may react and change on-the-fly how to reach its destination, selecting a route different from the original one.

To assess the social cost incurred by CA-CTA at the global level, we measure the moving average of the travel time, that is, how the average travel time of the entire population of drivers, computed over all the O-D pairs, evolves during the simulation. The results, with 95% confidence interval error bars, are plotted in Figure 7. In the beginning, the average travel time is similar for all the scenarios, but as the number of drivers that populate the network (i.e., its load) increases, it grows significantly faster with FCFS than with the CA-CTA policy. In terms of average travel times CTA is the best performing policy. CA-CTA has a slightly inferior performance, but it can be shown that it enforces an inverse relationship between bid value and delay, similar to the results presented in the previous section [27]. The fact that both CA-CTA and CTA outperform FCFS is an indication that, in general, a traffic assignment strategy (the "CTA" component of both policies) improves travel time. In fact,

FCFS drivers always select the shortest route, which in some cases is not the best route choice. Furthermore, granting reservations through an auction (the "CA" component of the CA-CTA policy) ensures that bid value and delay reduction are correlated.

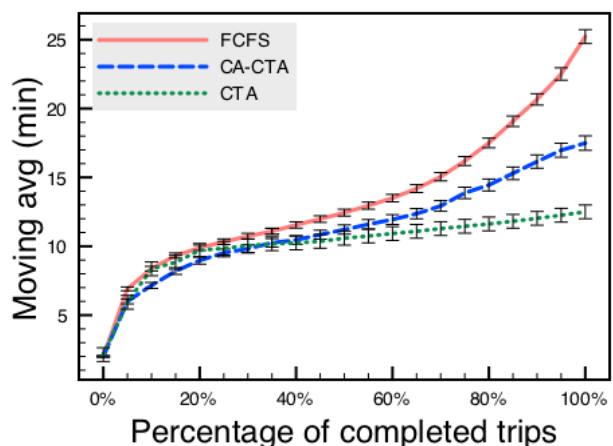

**Figure 7.** Moving average of travel times

*3.2. Evacuation Coordination in Smart Building*

The objective of an evacuation is to relocate evacuees from hazardous to safe areas while providing them with safe routes. Present building evacuation approaches are mostly static and preassigned (e.g. [30]). Frequently, no coordination is available except for predefined evacuation maps. Still, due to the lack of overall evacuation network information, there might be casualties caused by a too slow evacuation on hazardous routes. *Real-time* route guidance systems, which dynamically determine evacuation routes in inner spaces based on the imminent or ongoing emergency, can help reducing those risks. Chen and Feng in [31] propose two heuristic flow control algorithms for a real-time building evacuation with multiple narrow doors: with no limitation on the number of evacuation paths and k required evacuation paths, respectively. Filippoupolitis and Gelenbe in [32] proposed a distributed system for the computation of shortest evacuation routes in real-time. The routes are computed by decision nodes and are communicated to the evacuees located in their vicinity. However, this approach considers only the physical distance and the hazard present in each link and does not take into consideration crowd congestion on the routes.

A dynamic, *context-sensitive* notion of route *safety* is a key factor for such recommendations, in particular as herding and stampeding behaviours may occur at potential bottlenecks depending, among other factors, on the amount of people who intend to pass through them. Furthermore, smart devices allow guidance to be *personalized*, taking into account, for instance, the specific circumstance of the elderly, disabled persons, or families. In such settings, an adequate notion of *fairness* of evacuation route recommendations is of utmost importance to assure the trustworthiness of the system from the standpoint of its users [33]: the guidance should not only achieve good overall performance of the evacuation process, but must also generate proposals for each of its users that each of them perceive as efficient. Finally, large groups of people may need to be evacuated so *scalability* plays a key role.

Our proposal concentrates on real-time situation-aware evacuation guidance in smart buildings. The system aims at assigning efficient evacuation paths to individuals based on their mobility limitations, initial positions, respecting individual privacy, and other evacuation requirements. In our approach, a network of smart building agents calculate individual routes in a decentralized fashion. Complex event processing, semantic technologies and distributed optimization techniques are used to address this problem. In addition, we use the notion of agility to determine *robust* routes in the sense that they are not only fast but also allow finding acceptable alternatives in case of upcoming contingencies.

We rely on the existence of a rather extensive set of possible evacuation routes, which may be determined by evacuation experts or through some automated online or offline process. The different evacuation routes are stored in an emergency ontology that, together with an ontology describing the topological structure of the building specifies the a priori knowledge of our system. In addition, situational knowledge about the current situation in each moment of the building and of the evacuees is generated in real-time through a network of sensors. This dynamic knowledge is merged with the static knowledge about the infrastructure. In an emergency situation, semantic inference is used to select the most appropriate agile evacuation route for each individual in the building. Furthermore, real-time monitoring allows the system to reroute evacuees in case of contingencies and, thus, to propose evacuation routes that are adaptive to unpredictable safety drops in the evacuation network.

3.2.1. Distributed Architecture

The objective of the evacuation route guidance architecture (ERGA) is to provide individualized route guidance to evacuees over an app on their smartphones based on the evacuation information received from connected smartphones within the building and the building sensor network. However, even if an evacuee did not have a smartphone available, s/he could still receive information on relevant evacuation directions, e.g., through LED displays on the walls of a smart building.

ERGA (Figure 8) consists of user agents (UA) and a network of smart building (SB) agents.

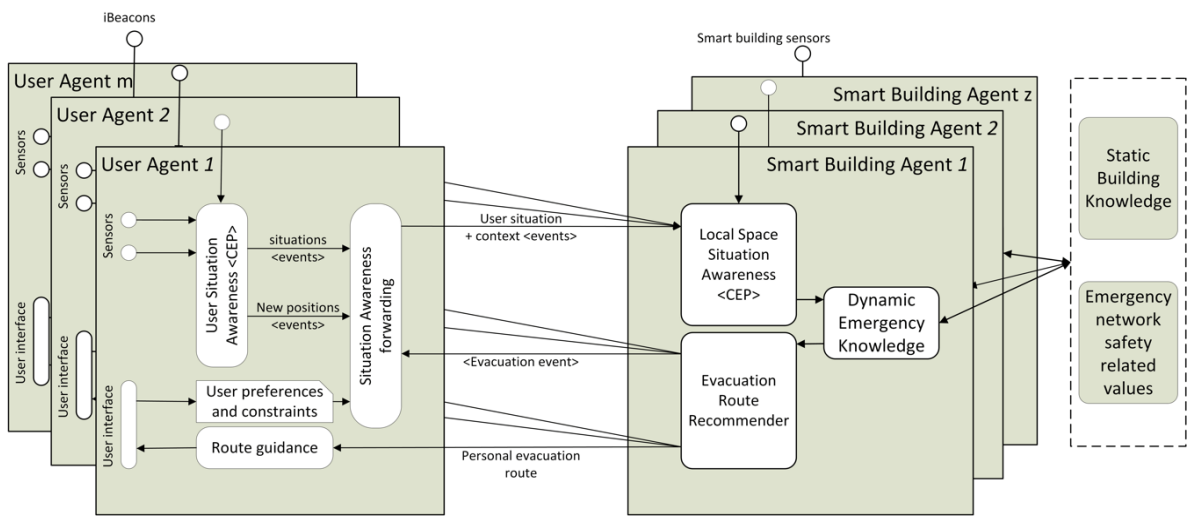

**Figure 8.** Situation-aware real-time distributed evacuation route guidance architecture (ERGA). User agents 1, 2, and m are located in the physical space of SB Agent 1 so that they are given route recommendations by SB Agent 1.

**User agents.** The user agent is associated with the application on a smartphone of an evacuee. It manages and stores all the information that is related to a specific evacuee in the building. Here, we assume that people that enter the building own a smartphone with the evacuation app installed, or they have been provided with some smartphone–like device that runs the app when they start to evacuate. The user agent contains three modules: (i) user preferences and constraints, (ii) user situation awareness module, and (iii) route guidance module.

The *user preferences and constraints* module allows defining constraints such as disabilities (e.g., the use of wheelchair or vision impairment) as well as evacuation–related behavioural disorders (e.g., agoraphobia, social phobia, etc.), while the preferences include the affiliate ties with other users of the building. The *user situation awareness* module exploits sensor data (from the smartphone and building) and reasons about the behaviour and location of the user. The presence of an evacuee together with the information derived from the situation awareness module and the individual preferences and constraints are passed to the closest SB agent. In order to assure privacy, only certain basic data about the user's situation should be forwarded to the SB agent (e.g., location, running events). In case of an emergency evacuation, the *user interface* provides the user with personalized navigation guidelines for evacuation.

**Smart building agents.** Situation awareness and decision making are distributed in the network of SB agents such that each agent is responsible of the semantic reasoning concerning the safety of its assigned physical space, as well as the evacuation route computation for the evacuees positioned in its physical space. We assume that each SB agent has at its disposal the information regarding all evacuation network's layout, topology and safety.

A single SB agent controls only its own physical space. The number and location of SB agents is defined when the system is installed. Each SB agent has a corresponding region (Voronoi cell) consisting of all user agents closer to that SB agent than to any other SB agent. Each SB agent contains a *local space situation awareness* module that perceives the safety conditions of the physical space it controls through combining and analysing the events provided by the sensors and individual user agents located within the smart space controlled by it. Moreover, each SB agent communicates with its neighbouring SB agents and with the user agents present within its physical space.

The *local space situation awareness* module functions in cycles. At the first phase, the local building sensor data is fused with the data from the locally present user agents. Then, the safety value is deduced. This data is sent to a blackboard or alike globally shared data structure containing the overall network safety values and visible to all agents. When an SB agent detects an emergency situation, it sends the updated safety value of its physical space to the shared blackboard. This allows, on the one hand, to monitor the real-time situation of the building and, on the other hand, to trigger an evacuation process and to execute control actions in such a process. SB agent's *evacuation route recommender* module computes optimized evacuation routes for each locally present user agent by distributed computation and communication with the rest of the SB agents in a multi-hop fashion. In this process, the algorithm uses: (i) data regarding the building topology, (ii) general knowledge about emergency and evacuation scenarios (e.g., facts that people with strong affiliate ties should always be evacuated together, the appropriateness of certain routes for people with limited mobility, and the influence of certain events like fire and smoke on the security level), and (iii) the current physical space situation awareness of the SB agent itself as well as regarding the evacuees that are currently in its space and evacuation network's safety values.

During evacuation, the global safety situation of the building is dynamically updated in real–time and each SB agent recalculates the evacuation routes if necessary.

3.2.2. Situation Awareness

We assume the existence of data provided by a smart infrastructure as well as by the users currently in the building. In particular, we require information for identifying the location of each user in the building.

There are various technological techniques to localize people in buildings. Measuring the strength of the signal of several WiFi access points could be used to calculate a person's location via trilateration. However, the signal strength is easily affected by the environment (obstacles, users, …) making it very difficult to obtain accurate positions. Another option is using RFID technology, but a lot of expensive readers would need to be installed in the building, and there are also similar trilateration problems as for WiFi. In addition, it would require providing an RFID tag to each person in the building. We opt for using Beacons, a recent technology to support indoor navigation. Beacons are cheap devices that emit Bluetooth signals, which can be read by beacon readers, in particular smartphones. Beacons send, among other information, a unique ID that allows identifying the specific sensor the user is near to, thus providing accurate user location.

Besides user location, other infrastructure sensors provide different measures such as temperature, smoke, fire, and so on. In addition, the users' smartphones built-in sensors provide information that allows detecting their activity (e.g. if the person carrying the phone is running).

Sensor events (each piece of information forwarded by or read from a sensor) are processed using Complex Event Processing (CEP), a software technology to extract the information value from event streams [34]. CEP analyses continuous streams of incoming events in order to identify the presence of complex sequences of events (event patterns). Event stream processing systems employ "sliding windows" and temporal operators to specify temporal relations between the events in the stream. The core part of CEP is a declarative event processing language (EPL) to express event processing rules. An event processing rule contains two parts: a condition part describing the requirements for firing the rule and an action part that is performed if the condition matches. An event processing engine analyses the stream of incoming events and executes the matching rules.

UAs exploit sensor data and infer the location and behaviour of their user. For example, data read from beacons is introduced as events of type *beaconEvent(beaconID)*. Then, the following CEP rule creates *enteringSection* and *leavingSection* events, meaning that the user is entering, respectively leaving a certain space. The rule describes the situation that a new *beaconEvent b2* has been read in the phone, where the beacon ID has changed. The symbol "->" indicates that event *b1* occurs before event *b2*.

>   *CONDITION:*
>     *beaconEvent AS b1 -> beaconEvent AS b2* ∧
>     *b1.id <> b2.id*
>   *ACTION:*
>     *CREATE enteringSection(userID, b2)*
>     *CREATE leavingSection(userID, b1)*

*enteringSection* and *leavingSection* events, as well as others like *runningEvent* (generated by a CEP rule that checks if the average velocity of the user is higher than 5 km/h for the last 10 seconds) are forwarded to the SB agent monitoring the user's location area.

SB agents receive processed events from the UA in their area. That information, as well as the obtained from smart building sensors, is incorporated into a stream of events. Again, the event stream is processed by the CEP engine generating more abstract and relevant situation awareness information. For instance, a panic event can be inferred if more than 40% of persons in a certain section of the building emit a running event.

Finally, situation awareness information, in form of events, is then transformed into a semantic representation, namely RDF facts. Afterwards, the situation information is ready to be consumed by semantic inference engines. We use OWL ontologies to represent information semantically in our system (user preferences, building topology, emergency knowledge, building situation). Semantic representations provide the means to easily obtain inferred knowledge. For example, if we define a class *DisabledPerson* to represent people with at least one disability, then we can infer disabled people even though they have not been explicitly described as instances of that class. For more complex reasoning tasks, we use rules on top of our OWL ontologies, which typically add new inferred knowledge. In particular, we use rules to determine the accessibility of certain sections in the building, and to select possible evacuation routes.

3.2.3. Personalized Route Recommendation

Our aim is to safely evacuate all the evacuees (or at least as many as possible) within an allotted upper time limit. This limit is usually given by the authorities in charge of evacuation.

Initially, we rely on the existence of a set of *predefined evacuation routes*. This set is independent of user constraints. The set of routes is analysed with the objective of generating *personalised efficient evacuation routes*, i.e. sets of alternative routes for each particular user considering the current situation of the building and user constraints (e.g. wheelchair, blind, kids, …). This is carried out in two steps. First, those routes that are not *time-efficient* (e.g. their expected evacuation time are not within the time limit) are filtered out. Next, using a rule-based system, *safe personalised plans* for each user are created. These routes only include traversing sections that are accessible for that particular user (e.g. avoiding paths through staircases if the person uses a wheelchair). Semantic rules and OWL reasoning are used in this task. For example, the following Jena[3] rule identifies staircase sections that are not accessible for people in a wheelchair:

> *(?user :hasDisability :Wheelchair)*
> *(?section rdf:type :Staircase)*
> *->*
> *(?section :notAccessibleFor ?user)*

The *personalised efficient evacuation routes* need to be ranked so as to select one route for each person in the building. We represent the evacuation network by a directed graph $G = (N, A)$, where $N$ is a set of $n$ nodes representing sections, and $A$ is the set of $m$ arcs $a = (i, j)$, $i, j \in N$ and $i \neq j$, representing walkways, doors, gateways, and passages connecting sections $i$ and $j$. Let $O \subseteq N$ and $D \subseteq N$ be a set of all evacuees' origins and safe exit destinations, respectively. We model the evacuation as a unified crowd flow, each individual is seen as a unit element (particle) of that flow and the objective is to maximize the flow of demands (evacuation requests) with certain constraints. We consider travel time optimization with path safety, envy-freeness (fairness) and agile paths.

**Route Safety.** Our objective is to safely evacuate as many evacuees as possible from all origins o ∈ O over the safest and the most efficient evacuation paths to any of the safe exits d ∈ D. Let us assume that safety status $S_a$ is given for each arc $a \in A$ as a function of safety conditions that can be jeopardized by a hazard. Safety can be calculated from sensor data (e.g. temperature, smoke…), and using space propagation models and aggregation functions to combine different influences and variables measured. A thorough description of this field can be found in, e.g., [35, 36]. We normalize it to the range [0, 1], such that 1 represents perfect conditions while 0 represents conditions impossible for survival, with a critical level for survival $0 < S^{cr} < 1$ depending on the combination of a the previously mentioned parameters.

If each constituent arc $a \in k$ of a generic path $k$ has safety $S_{a \in k} \geq S^{cr}$, then path $k$ is considered to be safe. On the contrary, a path is considered unsafe and its harmful effects may threaten the evacuees' lives. The proposed evacuation paths should all satisfy safety conditions $S^k \geq S^{cr}$. However, when such a path is not available, a path with the maximal safety should be proposed where the travel time passed in the safety jeopardized areas should be minimized. Since safety may vary throughout a path, we introduce a normalized path safety that balances the minimal and average arc safety values:

$$S^k = \sqrt[|a \in k|]{\prod_{a \in k} S_a}, \forall k \in P_o, o \in O$$

Where $P_o$ is the set of simple-paths from origin o ∈ O to an exit.

**Fair route recommendation.** An adequate notion of fairness of evacuation route recommendations is important to assure the trustworthiness of the system from the evacuees' viewpoint [33]: the guidance should not only achieve good overall performance of the evacuation process, but must also generate evacuation routes for each of the evacuees that each of them perceives as efficient and fair. For example, if there are two close-by evacuees at some building location, they should be proposed the same evacuation route, and if not possible, then the routes with similar safety conditions and evacuation time.

We aim at proposing available safe simple paths with a maximized safety acceptable in terms of duration in free flow for each evacuation origin. By acceptable in terms of duration in free flow, we mean the paths whose traversal time in free flow is within an upper bound in respect to the minimum free flow duration among all the available evacuation paths for that origin.

---

[3] jena.apache.org

The concept of envy-free paths is introduced in [37]. Basically, it defines a path allocation to be *α-envy-free* if there is no evacuee at origin *o'* that envies any other evacuee at origin *o* for getting assigned the path with a lower duration than $\alpha^{th}$ power of the path duration assigned to the evacuee on *o'*.

**Agile routes.** When an unpredicted hazard occurs on a part of the evacuation route, it becomes unsafe and impassable. If, in the computation of an evacuation route, we did not consider this fact and the related possibility to reroute to other efficient evacuation routes on its intermediate nodes, then, in case of contingency, re-routing towards safe areas might be impossible causing imminent fatalities of evacuees. A similar case may occur if, for example, a high flow of evacuees saturates an evacuation path and causes panic. Therefore, we prefer routes where each intermediate node has a sufficient number of dissimilar efficient evacuation paths towards safe exits, if possible within the maximum time of evacuation given for a specific emergency case. In that respect, evacuation centrality is defined in [38] as follows.

*Evacuation centrality* $C_\varepsilon(i)$ of node *i* is a parameter that represents the importance of node *i* for evacuation. The value of the evacuation centrality of the node is the number of available sufficiently dissimilar time-efficient evacuation paths from that node *i* towards safe exits.

Once when we find the evacuation centrality measure for each node of the graph, the objective is to find an evacuation path that maximizes the overall value of the intermediate nodes' centrality measures. We call every such path *agile evacuation path*; a path where an evacuee has higher chances to re-route in case of contingency in any of the intermediate nodes or arcs. *Path agility* $\Delta(k)$ is defined as:

$$\Delta(k) = \sqrt[|(i,j) \in k|]{\prod_{(i,j) \in k} C_\varepsilon(j)}$$

Since we are not concerned about the number of arcs in the path, we take the $|(i, j) \in k|^{th}$ root of the Nash product in this formula. We recommend the evacuation paths with the highest agility to evacuees and recompute this value every time the safety and/or congestion conditions change along the recommended path.

*3.3. Emergency medical service coordination*

The domain of medical assistance, includes many tasks that require flexible on-demand negotiation, initiation, coordination, information exchange and supervision among different involved entities (e.g., ambulances, emergency centres, hospitals, patients, physicians, etc.). In the case of medical urgencies, in addition, the need for fast assistance is evident. It is of crucial importance for obtaining efficient results, improving care and reduce mortality, especially in the case of severe injuries. Out of hospital assistance in medical urgencies is usually provided by Emergency Medical Assistance (EMA) services, using vehicles (typically ambulances) of different types to assist appearing patients at any location in a given area. In such services, the coordination of the available resources is a key factor in order to assist patients as fast as possible. The main goal here is to improve one of the key performance indicators: the response time (the time between a patient starts calling an EMA service centre and the moment medical staff, e.g., an ambulance, arrives at his location and the patient can receive medical assistance).

One way to reduce response times consists in reducing the part that depends on the logistic aspects of an EMA service through an effective coordination of the assistance vehicle fleet (for simplicity, here we assume a fleet of ambulances). In this regard, there are two principal problems EMA managers are faced with: the assignment or allocation of ambulances to patients and the location and redeployment of the ambulance fleet. The assignment or allocation problem consists in determining at each moment which ambulance should be sent to assist a given patient. And the location and redeployment consists in locating and possibly relocating available ambulances in the region of influence in a way that new patients can be assisted in the shortest time possible.

Most of recent works for coordinating ambulance fleets for EMA have been dedicated to the redeployment problem. A lot of work has concentrated on the dynamic location of ambulances, where methods are proposed to redeploy ambulances during the operation of a service in order to take into account the intrinsic dynamism of EMA services (e.g. [39, 40, 41]). Most proposals on dynamic redeployment of ambulances only consider the possibility to relocate ambulances among different, predefined sites (stations). This requirement is relaxed in the work proposed in [42], where a number of ambulances can be relocated to any place in the region. Regarding dispatching strategies (the patient allocation problem), most works use the "nearest available ambulance" rule for assigning ambulances to patients in a first-came first-served manner. Some works analyse priority dispatching strategies to account for severity level of patients ([42, 43]).

In our previous work [44], we have proposed a system that re-allocates ambulances to patients and redeploys available ambulances in a dynamic manner in order to reduce the average response times. Our redeployment approach differs from others in the sense that we do not try to maximize the zones in a region that are covered with respect to some time limits. Instead, we use an approach based on geometric optimization that tends to optimize in each moment the positions of all ambulances that are still available such that the expected arrival time to potential new emergency patients is minimized. With regard to the allocation of patients to ambulances, we propose a dynamic approach similar to [45] but, instead of optimizing the global travel times of all ambulances, we concentrate only on the sum of the arrival times of ambulances to the pending emergency patients. This system is summarised in this section.

We use the following notation to describe the problem and to present our solution. The set of ambulances of an EMA service is denoted by $A=\{a_1, \ldots, a_n\}$, where *n* is the cardinality of *A*. Even though, most EMA services employ different types of ambulances, for reasons of simplicity, we just consider a single type. Each ambulance has a position and an operational state which vary during time. $p(a_i)$ and $s(a_i)$ denote the current position and the current state of ambulance $a_i$, respectively. The position refers to a geographical location and the state can be one of the following:

- *assigned*: An ambulance that has been assigned to a patient and is moving to the patient's location.
- *occupied*: An ambulance that is occupied either attending a patient "in situ" or transferring him/her to a hospital.
- *idle*: An ambulance that has no mission in this moment.

We denote by $A^A$, $A^O$ and $A^I$ the sets of available, occupied and idle ambulances at a given moment.

Regarding the patients, $P=\{p_1, \ldots p_m\}$, denotes the current set of unattended patients in a given moment, e.g., patients that are waiting for an ambulance, where *m* is the cardinality of *P*. Each patient $p_j \in P$ has a location (denoted by $p(p_j)$). We assume that patients do not move while they are waiting for an ambulance, thus $p(p_j)$ is constant. Furthermore, once an ambulance has reached a patient's location in order to provide assistance, this patient is removed from *P*.

3.3.1. Dynamic re-assignment

The ambulance allocation problem consists in finding an assignment of (available) ambulances to the emergency patients that have to be attended. In current EMA services, mostly a priority dispatching strategy is used, where patients are assigned in a sequential order of appearance and patients with a higher severity level are assigned first. In each case, usually the nearest idle ambulance $a_i \in A^I$ is assigned. This can be seen as a first-call first-served (FCFS) rule, where patients with the same security level that called first are also assigned first to an ambulance. After an ambulance has been assigned to a patient, this assignment is usually fixed.

The FCFS approach is not always optimal from a global perspective. First, if more than one patients have to be attended it is not optimal in the sense that is does not minimize the response times to all patients. Furthermore, the dynamic nature of an EMA system implies that a given assignment of ambulances to patients at one point in time, might not be optimal at a later point, e.g., if new patients appear or an ambulance that has been occupied before is getting available again.

In order to reduce the average arrival time in the dynamic environment of an EMA service, the assignments of ambulances to patients could be optimized globally and the assignments should be recalculated whenever relevant events take place and a better solution may exist. Based on this idea, we proposed a dynamic assignment mechanism of ambulances to patients, which optimizes the assignments at a given point in time and recalculates optimal assignments when the situation changes.

Given a set of patients to be attended P and a set of ambulances that are not occupied, $A^A \cup A^I$, at a specific moment, the optimal assignment of ambulances to patients is a one-to-one relation between the sets $A^A \cup A^I$ and P, that is, a set of pairs AS={<$a_k,p_l$>,<$a_s,p_q$>,...} such that the ambulances and the patients are all distinct, and that fulfils the following conditions:

- The maximum possible number of patients is assigned to ambulances, that is:
  $\forall p_j \in P: \exists <a_i,p_j> \in AS$ if $n \geq m$ and $\forall a_i \in A^A \cup A^I: \exists <a_i,p_j> \in AS$ if $n<m$
- The total expected travel time of the ambulances to their assigned patients:
  $\sum_{<a_i,p_j> \in AS} ETT(p(a_i), p(p_j))$ is minimized

*ETT(x,y)* denotes the expected travel time for the fastest route from one geographical location *x* to another location *y*.

Calculating such an optimal assignment is a well-known problem which can be solved in cubic time, e.g., with the Hungarian method [46] or with Bertsekas' auction algorithm [47]. We propose to use the second approach because it has a naturally decentralized character and could be optimized in settings as the one analysed here.

An optimal assignment *AS* at a moment *t*, due to the dynamic nature of an EMA service, might become suboptimal at a time *t'* (*t'* > *t*). The following cases need to be considered:

1. One or more new patients require assistance: In this case, the set of patients that have to be attended changes and the current assignment *AS* may not be optimal any more.
2. Some ambulances that have been occupied at time *t* have finished their mission and are idle at time *t'*. These ambulances could eventually improve the current assignment.

Based on this analysis, we propose a dynamic system based on an event-driven architecture and that recalculates the global assignment whenever one of the following events occur: *newPatient($p_j$)* (a new patient has entered the set *P*) or *ambFinishedEvent($a_i$)* (an ambulance that was occupied before, is getting idle again). In the recalculation of an existing assignment, ambulances that have been already dispatched to a patient, but did not reach the patient yet, may be de-assigned form their patients or might be re-assigned to other patients. This approach assures that the assignment *AS* is optimal, with regard to the average travel time to the existing patients, at any point in time.

3.3.2. Dynamic re-deployment

The second part of the proposed coordination approach for EMS services consists in locating and redeploying ambulances in an appropriate manner. Here, the objective is to place ambulances in such a way that the expected travel time to future emergency patients is minimized.

We address this problem by using Voronoi tessellations [48]. A Voronoi tessellation (or Voronoi diagram) is a partition of a space into a number of regions based on a set of generation points, and such that for each generation points there will be a corresponding region. Each region consists of the points in the space that are closer to the corresponding generation point than to any other. Formally, let $\Omega \in R^2$ denote a bounded, two-dimensional space and let $S=\{s_1,...,s_g\}$ denote a set of generation points in $\Omega$. For simplicity, let $\Omega$ be a discrete space. The Voronoi region $V_i$ corresponding to point $s_i$ is defined by:

$V_i = \{y \in \Omega : |y - s_i| < |y - s_j| \text{ for } j=1,...,k \text{ and } j \neq i\}$

where $|\cdot|$ denotes the Euclidean norm. The set $V(S)=\{V_1, ..., V_k\}$ with $\bigcup_{i=1}^{k} V_i = \Omega$ is called a Voronoi tessellation of *S* in $\Omega$. A particular type of tessellations are Centroidal Voronoi Tesselation (CVT). A centroidal Voronoi tessellation is one where each generation point $s_i$ is located in the mass centroid of its Voronoi region w.r.t. some positive density function $\rho$ on $\Omega$. A CVT is a necessary condition for minimizing the cost function and, thus, provides a local minimum for the following cost function:

$$F(S) = \sum_{V_i \in V(S)} \sum_{y \in V_i} \rho(y) \cdot |y - s_i|^2 \tag{1}$$

A common approach to calculate CVTs and, thus, to minimize (1) is the algorithm proposed by Lloyd [49]. The algorithm is an iterative gradient descent method that finds a set of points *S* that minimizes *F(S)* in each iteration and converges to a local optimum. Lloyd's algorithm can be summarized in the following steps:

1. Select an initial set *S* of *k* sites in *Ω*
2. Calculate the Voronoi regions $V_i$ for all generation points $s_i \in S$
3. Compute the mass centroid of each region $V_i$ w.r.t. the density function *q*. These centroids compose the new set of points *S*.
4. If some termination criteria is fulfilled, finish; otherwise return to step 2.

Lloyd's algorithm is not assured to find a global minimum, but it finds solutions of reasonable quality very fast – after a few iterations. Therefore, we apply Lloyd's algorithm to find suboptimal positions of ambulances. The application is straight forward. *Ω* represents the region of interest and the set of generation points *S* represents the set of all idle ambulances and their positions. Each Voronoi region $V_i$ corresponds to the area that is covered by ambulance $s_i$, e.g., the area for which $s_i$ is the closest ambulance in case any patient requires help. Furthermore, as the density function $\rho$, we use an estimation of the probability distribution of the appearance of a patient in any point in the region of interest. In particular, to define $\rho$, we divide the region of interest in a set of equally sized cells $C=\{c_1, \ldots, c_u\}$, where *u* is the cardinality of *C*. That means we discretize the region *Ω* into *u* points, each of which represents the centre of one of the cells $c_i$. Then, we estimate for each cell $c_i$ the conditional probability $p_{ci}$ that a new emergency patient will appear in this cell. With this setting, Lloyd's algorithm finds a distribution of ambulances that minimizes (1). In particular, in our case this is a reasonable approximation for minimizing the expected distance and, thus, the arrival time, to future emergency patients.

The probabilities $p_{ci}$ can be obtained by tracking historical data on emergency cases. Furthermore, it is possible to model different situations (like seasons, day of the weak, time interval, etc.) through different probability distributions.

In the application of Lloyd's algorithm, at a given point in time, we use the positions of all idle ambulances as the set of initial generation points *S*. We apply a fixed number of iterations (in the experiments we use 50) and the resulting set *S* represents the "recommended" distribution of ambulances at this particular moment. The new positions are sent to the ambulances and the ambulances such that they can move towards that positions. Because of road and parking conditions, the obtained positions are rather considered as indications of an area. That is, once given such an area, the ambulance driver will decide which is the most appropriate waiting location in that area.

We use the Euclidean norm as a distance measure to generate the Voronoi regions for the ambulances. While in a real traffic scenario, as it is our case, the Euclidean distance is only a rather imprecise approximation of real distances on the road network, from a global perspective, and assuming a rather homogeneous connection between different points of the region of interest (as it is usually the case in many urban areas), the Euclidean norm seems to work reasonably well for our purposes. Furthermore, using "road-network distances" when calculating the Voronoi regions is a rather complicated task that would increase the computation complexity considerably.

Similar to the ambulance assignment problem, the dynamic nature of an EMA service implies that the optimal positions of the idle ambulances will change when changes in the environment occur. In order to cope with such changes, the ambulance positions are recalculated dynamically whenever any of the following events occur:

- An ambulance that was assigned to a patient has been de-assigned.
- An ambulance has finished a patient assistance mission and has changed its state from occupied to idle.
- An ambulance that was idle has been requested to assist a patient. It changes its state from idle to assigned.
- The situation that determines the underlying probability distribution changes (e.g., a distribution for a new time interval day of the week or season should be used)

3.3.3. Experimental results

We tested the effectiveness of the dynamic re-assignment and re-deployment approaches in different experiments simulating the operation of SUMMA112, the EMA service provider organization in the Autonomous Region of Madrid in Spain. We used a simulation tool that allows for a semi-realistic simulation of intervals of times of normal operation of an EMA service. The tool reproduces the whole process of attending emergency patients, from their appearance and communication with the emergency centre, the schedule of an ambulance, the "in situ" attendance and, finally, the transfer of the patients to hospitals. The simulator operates in a synchronized manner based on a step-wise execution, with a step frequency of 5 seconds. That is, every 5 seconds, the activities of all agents are reproduced leading to a new global state of the system. In the simulations we are mainly interested in analysing the movements of ambulances and the subsequent arrival times to the patients. The movements are simulated using an external route service to reproduce semi-realistic movements on the actual road network with a velocity adapted to the type of road. External factors, like traffic conditions or others, are ignored. The duration of the phone call between a patient and the emergency centre and the attendance time "in situ" are set to 2 and 15 minutes, respectively. As the area of consideration, we used a rectangle of 125×133 km that covers the whole region of Madrid. For calculating the probability distribution of upcoming patients, we divided the region in cells with size 1300×1300 meters. A different probability distribution is estimated for each day of the week and each hour from statistical data (patient data from the whole year 2009). We used 29 hospitals (all located at their real positions) and 29 ambulances with advanced life support (as it was used by SUMMA112 in 2009) and we simulated the operation of the service for 10 different days (24 h periods) with patient data from 2009 (in total 1609 patients). The days where chosen to have a representation of high, medium and low workloads. We only take into account so called level 0 patients, e.g., patients with a live threading situation.

We compare two approaches:
- SUMMA112: the classical approach (used by SUMMA112). Patients are assigned to the closest ambulances using a fixed FCFS strategy. Furthermore, ambulances are positioned on fixed stations (at the hospitals), waiting for missions. After finishing any mission, the ambulances return to their station.
- DRARD: In this case, the dynamic re-assignment and re-deployment methods are employed. With regard to dynamic re-employment, idle ambulances only move to a new recommended position if it is further away than 500 meters. This is to avoid short, continuous movements.

Table 1 presents the average arrival times (in minutes) obtained with the two models in simulations for the 10 selected days. As the results show, the use of the DRARD approach provides a considerable improvement (between around 10 and 20%). If we look on all 1609 patients, the average times are 11:45 and 9:54 minutes, respectively, which implies an improvement of 15.8%.

**Table 1.** Average arrival times in minutes for 10 different days.

| Day/ #patients | 1/221 | 2/152 | 3/199 | 4/124 | 5/137 |
|---|---|---|---|---|---|
| SUMMA | 11:42 | 11:52 | 11:03 | 11:23 | 11:50 |
| DRARD | 9:46 | 9:50 | 9:29 | 9:39 | 10:09 |
| Improvement % | 16.6 | 15.7 | 12.8 | 9.3 | 14.0 |
| Day/ #patients | 6/175 | 7/96 | 8/160 | 9/144 | 10/201 |
| SUMMA | 12:30 | 12:51 | 12:42 | 10:11 | 11:49 |
| DRARD | 10.51 | 9:48 | 10:05 | 9:05 | 10:08 |
| Improvement % | 13.1 | 23.7 | 20.7 | 10.8 | 14.3 |

In Figure 9 we present the results of the distribution of arrival times for the different approaches for all 1609 patients of the 10 selected days. The patients in each curve are ordered by increasing arrival time. A clear difference can be observed between the DRARD method with respect to the current operation model of SUMMA112. The results are clearly better for almost all arrival time ranges. Furthermore, the most important improvements can be observed in the range of higher times. This is a very positive effect because it assures that more patients can be attended within given response time objectives. For example, out of the 1609 patients, 1163 (72.3%) are reached within 14min with in the SUMMA model, whereas this number increases to 1356 patients (84.3 %) with DRARD.

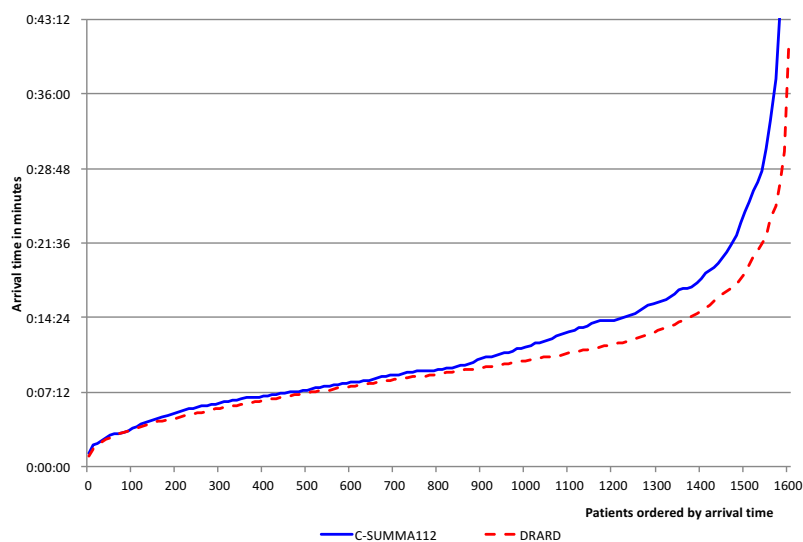

**Figure 9.** Comparison of arrival times on patients. Here, the 1609 patients from the 10 analysed days are ordered with respect to the arrival time in each curve.

As shown in the results, the proposed dynamic re-assignment and re-deployment methods clearly improve the efficiency of a EMA service in terms of reducing response time. However, the approaches, in particular dynamic re-deployment introduce an extra cost. Since the mechanism is based on an almost continuous repositioning of idle ambulances, the travel distances the ambulances have to do increase. Considering the 10 days, the average distance each ambulance has to cover each day in the SUMMA model is 95.48 km, whereas it is 299,97 km for the DRARD approach. This is, ambulances have to travel about three times the distance because of frequent location changes. It is a political decision whether this extra effort is acceptable in order to improve the quality of service. In any case, compared to augmenting the number of ambulances in order to reduce response times, the DRARD approach appears to be a less costly alternative. In this sense, we have executed the DRARD method also with less ambulances and roughly the same average arrival time than obtained with 29 ambulances in the SUMMA model, can be obtained with 21 ambulances and the DRARD approach.

*3.4. Distributed coordination of emergency medical service for angioplasty patients*

Based on the World Health Organization data, ischemic heart disease (IHD) is the single most frequent cause of death killing 8.76 million people in 2015, and one of the leading causes of death globally in the last 15 years [50]. It is a disease characterized by ischaemia (reduced blood supply) of the heart muscle, usually due to coronary artery disease. At any stage of coronary artery disease, the acute rupture of an atheromatous plaque may lead to an acute myocardial infarction (AMI), also called a heart attack. AMI can be classified into acute myocardial infarction with ST-segment elevations (STEMI) and without ST elevation (NSTEMI). Effective and rapid coronary reperfusion is the most important goal in the treatment of patients with STEMI.

One of the reperfusion methods is angioplasty or primary percutaneous coronary intervention (PCI). It is the preferred treatment when feasible and when performed within 90 minutes after the first medical contact [51,52]. Due to insufficient EMS coordination and organizational issues, elevated patient delay time, defined as the period from the onset of STEMI symptoms to the provision of reperfusion therapy, remains a major reason why angioplasty has not become the definitive treatment in many hospitals.

Conventional EMS procedure in assisting AMI emergencies is the following. Patients are diagnosed in the place where they suffer chest pain: at their momentary out-of-hospital location or at a health centre without angioplasty. In both cases the ECC applies First-Come-First-Served (FCFS) strategy and locates the nearest available (idle) ambulance with Advanced Life Support (ALS) and dispatches it to pick up the patient. After the ambulance arrives to the scene and diagnoses AMI by an electrocardiogram, ambulance confirms the diagnosis to the ECC which has a real time information of the states of ambulances. The ECC applies FCFS strategy for hospital and cardiology team assignment by locating the nearest available hospital with catheterization laboratory and alerting the closest hospital cardiology team of the same hospital.

The improvements of the EMS coordination in the literature are achieved both by novel fleet real-time optimization and communication methods, as by new multiagent models, see, e.g., [¡**Error! No se encuentra el origen de la referencia.**, 52, 54, 55]. Despite of an exhaustive quantity of work on the optimization of EMA, to the best of our knowledge, there is little work on optimization models for the coordination of EMS when the arrival of multiple EMS actors needs to be coordinated for the beginning of the patients' treatment. This is the case with STEMI patients assigned for angioplasty treatment where, in the case of multiple angioplasty patients, the FCFS strategy discriminates the patients appearing later.

EMA coordination for STEMI patients includes the assignment of three groups of actors: assignment of idle ambulances to patients, assignment of catheterization laboratories in available hospitals to patients receiving assistance in-situ, and assignment of available cardiology teams to hospitals for the angioplasty procedure performance. All of the three assignments need to be combined in a region of interest such that the shortest arrival times are guaranteed to all patients awaiting angioplasty at the same time. In continuation, we present the solution approach from [52], which presents a coordination model for EMS participants for the assistance of angioplasty patients. The proposed approach is also applicable to emergency patients of any pathology needing pre-hospital acute medical care and urgent hospital treatment.

We concentrate on the minimization of the patient delay intended as the time passed from the moment the patient contacts the medical emergency coordination centre (ECC) to the moment patient starts reperfusion therapy in the hospital. The patient delay defined in this way is made of the following parts, Figure 10:

T1    Emergency call response and decision making for the assignment of EMS resources;
T2    Mobilization of an idle ambulance and its transit from its current position to the patient;
T3    Patient assistance in situ by ambulance staff;
T4    Patient transport in the ambulance to assigned hospital;
T5    Cardiology team transport from its momentary out-of-hospital position to the hospital;
T6    Expected waiting time due to previous patients in the catheterization laboratory (if any).

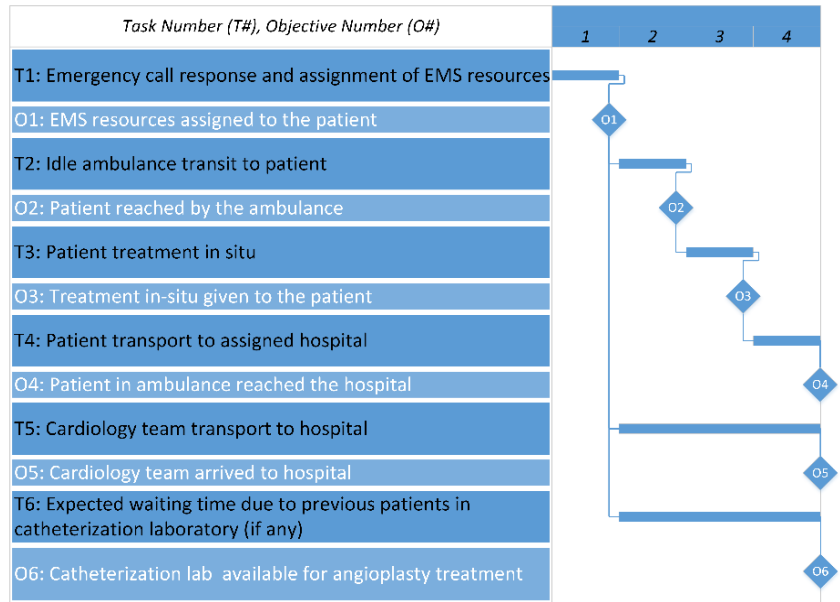

**Figure 10.** Gantt diagram of the coordination of EMS for angioplasty treatment

The optimal patient delay time for a single patient is the lowest among the highest values of the following three times for all available ambulances and angioplasty-enabled hospitals, Figure 10:

- the expected patient delay time to hospital (the sum of times T2, T3, and T4, in continuation represented by parameters *t(a,p)*, *t(p)*, and *t(p,h)*, respectively.
- the expected minimal arrival time among cardiology teams to the same hospital (T5), represented by $min_{c \in C_{av}} t(c,h)$
- and the expected shortest waiting time until hospital *h* gets free for patient *p*, $min\ \rho_{h,p}$ (T6).

For simplicity, we let $t_{php} = max_{h \in H_{av}}(t(p,h), min\ \rho_{h,p})$ for all patients $p \in P$. Then, from the global point of view, considering all pending out-of-hospital patients, the problem transforms into:

$$min \Delta t_P = \sum_{p \in P} \Delta t_P = \sum_{p \in P} t(a,p) + \sum_{p \in P} t(p) + \sum_{p \in P} \left( max_{h \in H_{av}} \left( t_{php}, min_{c \in C_{av}} t(c,h) \right) \right) \quad (1)$$

subject to

$$\Delta t_p \leq t_p^{max}, \forall p \in P. \quad (2)$$

The overall patient delay time $\Delta t_P$ in Figure 10 is an additive function. Since the minimum arrival times cannot be always guaranteed for all patients due to the limited number of EMS resources, a sum of the EMA tasks' durations should be minimized for each patient individually and for the system globally considering individual constraints. This gives an underlying linear programming structure to the EMS coordination problem. Therefore, it is possible to guarantee optimal outcomes even when the optimization is performed separately on individual sum components, i.e., when ambulance assignments are negotiated separately from the hospital and cardiology team assignment, e.g., [56, 57]. This fact significantly facilitates the multiagent system's distribution and enables a multi-level optimization. Hence, we decompose optimization problem (1)-(2) as follows. On the first level, we assign ambulances to patients such that the expected arrival time of ambulances to patients *t(a,p)* is minimized. Note that since *t(p)* in (1) is a constant for every patient *p* depending only on the patient's pathology and not on the assigned ambulance, we can exclude it from the optimization.

Then, on the second optimization level, we approach the second part of (1) $\sum_{p \in P}\left(max_{h \in H_{av}}\left(t_{php}, min_{c \in C_{av}} t(c,h)\right)\right)$ which is an NP-hard combinatorial problem. However, by approximating (1) with a sequence of problems where we first decide on the assignment of hospitals to pending patients and then assign cardiologists to patients already assigned to hospitals, we obtain two linear programs to which we can apply tractable optimal solution approaches such as the auction algorithm [47]. By decomposing (1) as can be seen in Figure 11 and allowing for reassignment of resources based on the adaptation to contingencies in real time, we obtain a flexible EMS coordination solution.

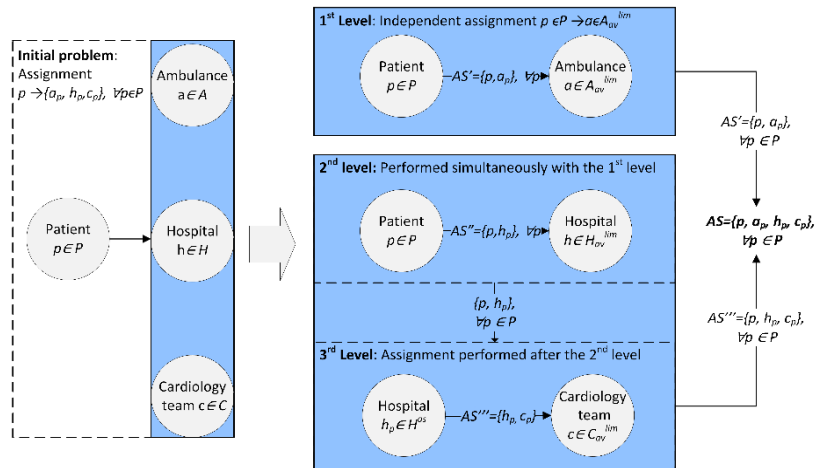

**Figure 11.** Proposed Three-level decomposition of the problem of EMS coordination for STEMI patients awaiting angioplasty.

In the following, we propose a change of the centralized hierarchy-oriented organizational structure to a patient-oriented distributed organizational structure of EMS that increases the flexibility, scalability, and the responsiveness of the EMS system. The proposed decision-support system is based on the integration and coordination of all the phases EMS participants go through in the process of emergency medical assistance (EMA). The model takes into consideration the positions of ambulances, patients, hospitals, and cardiology teams for real-time assignment of patients to the EMS resources.

3.4.1. EMA for angioplasty patients

The emergency medical system for the assistance of patients with STEMI is made of the following participants: patients, hospitals with angioplasty facilities, Medical Emergency Coordination Centre (ECC), ambulances staff, and cardiology teams, each one compound of a cardiologist and one or more nurses.

Usually, each hospital with angioplasty has assigned to it its own cardiology team(s) positioned at alert outside the hospital and obliged to come to the hospital in the case of emergency. This is because the cardiology teams' costs make a large portion of the overall costs in surgical services [55].

The objective of the proposed system is the reduction of patient delay times by distributed real-time optimization of decision-making processes. In more detail, we model patient delay time and present a three-level problem decomposition for the minimization of combined arrival times of multiple EMS actors necessary for angioplasty. For the three decomposition levels, we propose a distributed EMS coordination approach and modify the auction algorithm proposed by Bertsekas in [47] for the specific case. The latter is a distributed relaxation method that finds an optimal solution to the assignment problem.

On the first level, agents representing ambulances find in a distributed way the patient assignment that minimizes arrival times of available ambulances to patients. After the treatment in situ, on the second optimization level, ambulances carrying patients are assigned to available hospitals. On the third level, arrival times of cardiology teams to hospitals are coordinated with the arrival times of patients. The proposed approach is based on a global view, not concentrating only on minimizing single patient delay time, but obtaining the EMS system's best solution with respect to the (temporal and spatial) distribution of patients in a region of interest.

3.4.2. Simulation experiments

In this Section, we describe settings, experiments, and results of the simulated emergency scenarios that demonstrate the efficiency of the coordination procedure and a significant reduction in the average patient delay. We test the proposed approach for the coordination of EMS resources in angioplasty patients' assistance focusing on the average patient delay time in the case of multiple pending patients. We compare the performance of our approach with the FCFS method since it is applied by most of the medical emergency-coordination centres worldwide.

To demonstrate the scalability of our solution and its potential application to small, medium and large cities and regions, in the experiments, we vary the number of EMA ambulances from 5 to 100 with increment 5 and the number of angioplasty-capable hospitals from 2 to 50 with increment 2. The number of cardiology teams |C| in each experiment equals the number of hospitals |H|. Thus, the number of setup configurations used, combining different numbers of ambulances and hospitals with cardiology teams, sums up to 500.

For each configuration, we perform the simulation on 3 different instances of random EMS participants' positions since we want to simulate a sufficiently general setting applicable to any urban area that does not represent any region in particular. The EMS participants are distributed across the environment whose dimensions are 50×50 km. In each instance, we model hospital positions and the initial positions of ambulances, out-of-hospital cardiology teams, and patients based on a continuous uniform distribution. Therefore, each configuration can be considered as a unique virtual city with its EMS system. Assuming that the EMS system is placed in a highly dense urban area, this kind of modelling of the positions of EMS participants represents a general enough real case since the election of the hospital positions in urban areas is usually the result of a series of decisions developing over time with certain stochasticity, influenced by multiple political and demographical factors.

In the simulations, ambulances are initially assigned to the base stations in the hospitals of the region of interest. Additionally, we assume that after transferring a patient to the hospital, an ambulance is redirected to the base station where it waits for the next patient assignment. Furthermore, we assume that the hospitals have at their disposal a sufficient number of catheterization laboratories so that the only optimization factor from the hospital point of view is the number of available cardiology teams. If there are more patients with the same urgency already assigned waiting for treatment in the same hospital, they are put in a queue.

The simulation of each instance is run over a temporal horizon in which new patients are generated based on a certain appearance frequency. The EMS resources are dynamically coordinated from the appearance of a patient until the time s/he is assisted in hospital by a cardiology team. Each instance simulation is run over the total of 300 patients whose appearance is distributed equally along the overall time horizon based on the following two predetermined frequency scenarios: low (1 new patient every 10 time periods) and high (1 new patient every 2 time periods).

The period between two consecutive executions of the EMS coordination algorithm is considered here as a minimum time interval in which the assignment decisions are made; usually it ranges from 1 to 15 minutes. In each period, the actual state of EMS resources and pending patients is detected and the EMS coordination is performed such that the EMS resources are (re)assigned for all patients. To achieve an efficient dynamic reassignment of ambulances, the execution of the EMS coordination algorithm is furthermore performed with every new significant event, i.e., any time there is a significant change in the system due to new patients, or the significant change of travel time or state of any of the EMS participants.

In the experiments, we test the performance of the proposed EMS coordination method with respect to the FCFS benchmark approach. The comparison is based on the relative performance function $P = (t_{FCFS} - t_{OR})/ t_{FCFS} \cdot 100$ [%], where $t_{FCFS}$ and $t_{OR}$ are average patient delay times of the benchmark FCFS approach and the proposed model, respectively.

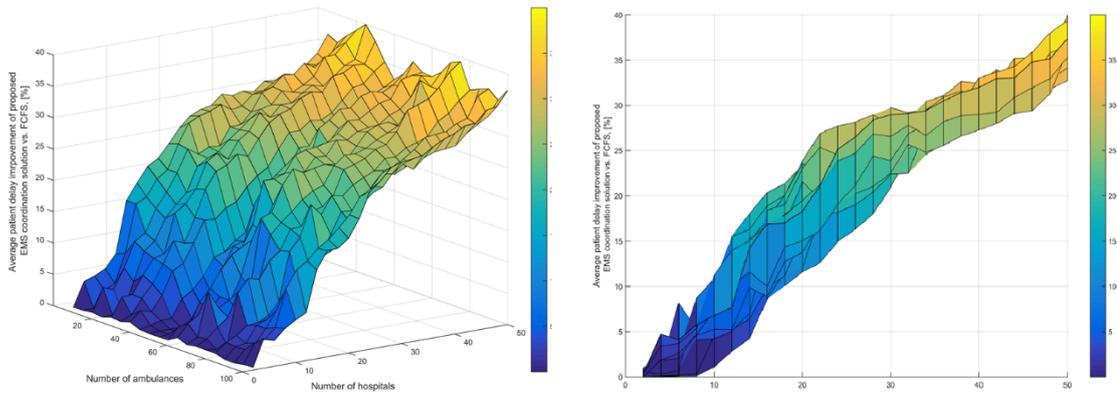

**Figure 12.** Avg. patient delay time of the proposed EMS coordination approach vs. the FCFS strategy [%] for the frequency of appearance 1 patient each 10 time periods

The simulation results of performance function P for the two simulated cases of patient frequency appearance of 1 and 5 new patients every 10 time periods are presented in the following. The performance of the proposed approach increases as the number of angioplasty enabled hospitals increases from almost identical average patient delay in the configuration setup with 2 hospitals up to 87,14 % with 50 hospitals, as can be seen in Figure 13. Observing the performance dynamics with respect to the varying number of hospitals, it is evident that the performance of the proposed EMS coordination method increases on average proportionally to the increase of the number of hospitals.

With a relatively low number of angioplasty-enabled hospitals (less than 15), our proposed EMS coordination approach performs on average better than FCFS up to 15 %. As the number of hospitals increases, the performance improves on average up to the maximum of 39,98 % for the first case, Figure 12 and up to 87,14 % , for the second case, Figure 13. However, mean patient delay improvement for the two cases is 35 % and 45.5 % respectively.

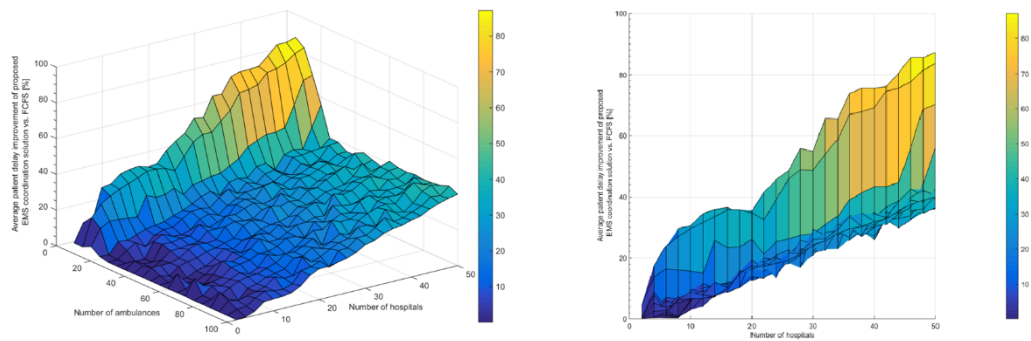

**Figure 13.** Avg. patient delay time of the proposed EMS coordination approach vs. the FCFS strategy, [%] for the frequency of appearance of 1 patient each 2 time periods

The static assignment of the FCFS principle discriminates against patients appearing later. Since ambulances are not equally distributed in the area, the proposed EMS coordination method compensates for the lack of EMS resources and their unequal distribution by reassigning them dynamically to pending patients. Dynamically optimized reassignment of EMS resources in real time is the main key to the improvement of the system's performance.

Thus, proportional to the increase of the number of hospitals, there is a constant improvement of performance. Even though the velocity of the EMS actors is not a relevant factor in the comparison of the performance of our proposed EMS coordination solution and the FCFS method, looking individually at the performance of each one of these methods, it is evident that the assignment cost accumulated through the time will be lower when the velocity of the EMS actors is higher.

Our simulation results show the efficiency of the proposed solution approach, resulting in significantly lower delay times for angioplasty on average. Of course, the effectiveness of the proposed model depends on the initial classification of patients, and the related determination of the urgency of their cases, as well as on the timely availability of cardiology teams and hospitals. Still, as the current experience shows, good quality patient assessments and the EMS resource availabilities can be assured in practice.

To implement our approach in practice, a patient's location needs to be known to the system. Ideally, patients should contact the ECC through a mobile phone with GPS for easier location. In addition, ambulances should have a GPS and a navigator for localizing the patient and navigating the way to him/her, as well as a means of communication with the rest of the EMS participants, and a digitalized map showing ambulances, patients and hospitals.

Moreover, hospitals should have a digitalized receptionist service to receive and process relevant data of a patient before his/her arrival. None of these requirements go significantly beyond the current state of affairs in major cities (such as Madrid). Moreover, there are intrinsic uncertainties present in the EMS coordination. In our experiments, we assume that travel times can be accurately forecasted, which, of course, is an important factor for the performance of the proposed system. In reality, this may not always be the case, as real world traffic conditions are notoriously hard to predict. However, there is abundant literature on traffic-aware vehicle route guidance systems tackling this problem, and we believe that such systems can be easily integrated into our approach. Still, an effective proof of this conjecture is left to future work.

*3.5. Coordination of transportation fleets of autonomous drivers*

A similar problem to the emergency medical service coordination consists in coordinating fleets for transportation in an urban area, e.g., messaging services or taxi fleets. However, in contrast to medical emergency services, here the objective is not only focused on response time, but also on cost efficiency. Furthermore, a primary characteristic of such systems, at least with the boom of collaborative economy, is that such types of fleets may be open [58] in the sense that private persons may participate in the fleet as autonomous workers, with their own vehicle and at different time intervals. With regard to the coordination of such fleets, the autonomy of the drivers is a crucial characteristic. It implies that the drivers get their income on a per-service basis instead of a monthly salary. This means that, besides accepting a set of basic rules, drivers are more concerned with the actual service they have been assigned from the system and may have more freedom to accept or decline assignment decisions.

Maciejewski et al. [59] present a real-time dispatching strategy based on solving the optimal taxi assignment among idle taxis and pending requests at certain intervals or whenever new events (new customer/available taxi) take place. Zhu and Prabhakar [60] analyse how suboptimal individual decisions lead to global inefficiencies. While most existing approaches try to minimize the average waiting time of customers, other works have a different focus. BAMOTR [61] provides a mechanism for fair assignment of drivers, i.e. to minimize the differences in income among the taxi drivers. For that, they minimize a combination of taxi income and extra waiting time. Gao et al. [62] propose an optimal multi-taxi dispatching method with a utility function that combines the total net profits of taxis and waiting time of passengers. Meghjani, and Marczuk [63] propose a hybrid path search for fast, efficient and reliable assignment to minimize the total travel cost with a limited knowledge of the network. In contrast to the previous works, the main characteristic of our approach is the possibility of modifying the assignment when a taxi has been dispatched but has not yet picked up a customer. In this sense, we followed a similar approach to the emergency medical service coordination presented in section 3.3. One of the few other works in this line is [64], which presented an adaptive scheduling in which reassignment is possible during a time interval until pick-up order is sent to the taxi and customer. In our case, we do not restrict reassignment to a specific interval. Furthermore, we propose a method that economically compensates the negatively affected taxis in the new schedule such that they do not have a loss in their income.

We consider a system that uses some mediator in charge of matching the transportation requests with vehicles of the fleet. On one hand, customers contact the mediator via some telematics means, requesting a transportation service. On the other hand, drivers subscribe to the system offering their services during specific time intervals where they are available.

We assume a payment structure for transportation services where clients pay a fixed price plus a fare per distance, as described in the following:

- From the client side, the price of a transportation service *s* is determined by the distance of the requested service plus a fixed cost:

$$Price(s) = fcost + fare \cdot d(s),$$

Where *d(s)* denotes the distance from the service origin to its destination, *fcost* is a fixed cost, *fare* is a rate a client has to pay per distance unit.

- From the driver or vehicle side, for a driver *v* the earnings depend on the price the client pays minus the cost of the vehicle for traveling the distance from the current position of v towards the position $p_o(s)$ and later to the destination *d(s)*:

$$Earn(v, s) = Price(s) - vcost \cdot ( d(v,s)+d(s) )$$

Where *d(v,s)* denotes the distance from the current position of the vehicle in the moment of assigning a service *s* to the origin point of that service and *vcost* is the actual cost rate of moving the vehicle on a per distance basis. *vcost* will implicitly include petrol, maintenance, depreciation of the vehicle, etc.

For simplicity, here we assume that *fcost*, *fare* and *vcost* are the same for all services and vehicles, that is, in the price structure we do not distinguish between different vehicles costs nor between different requests. If the use of different cost factors is important, the proposed model could be adapted accordingly. Furthermore, part or all of the amount of *fcost* could be retained by the mediator service as income. In this case, the earnings of the driver would be reduced by this amount.

Like in the emergency medical case, a typical approach for assigning services to vehicles in such a system is the first-call/first-served (FCFS) rule where each incoming request is assigned to the closest available driver in that moment and no re-assignments are done. As shown in [59], if there are more unassigned requests than available vehicles, it is better to assign vehicles to service requests. We call this strategy nearest-vehicle/nearest-request (NVNR).

Dynamic assignment strategies can improve the overall efficiency of a fleet. We assume that drivers, once they are available, are obliged to accept a transportation service assigned to them. However, in contrast to the emergency medical scenario presented in the previous section, drivers do not have to accept changes in the assignments. That means, they are free to accept or to decline proposed re-assignments from the system, considering their own objectives and benefit. So, the dynamic reassignment approach, as proposed in section 3.3.1 cannot be applied directly, as a driver would not be willing to accept a re-assignment that reduces his/her net income. In order to still taker advantage from cost reduction through re-assignment, we developed an incentive schema so as to convince drivers to accept re-assignments that are economically efficient from the global perspective of the system. The approach is detailed in the next subsection.

3.5.1. Dynamic re-assignment with compensations

The idea of coordinating the assignments of transportation tasks to (autonomous) drivers is similar to the one proposed in section 3.3.1: we want to reduce globally the total travel distance (and proportionally time) towards the origin points of the requested services. However, due to the rules of the system, a driver will usually not accept a "worse" task than the one he is already assigned to. We assume drivers to be economically rational, that is, they want to maximize their net income and minimize the time spent on their trips. In particular, with make the following assumptions:

- A driver would always prefer a task with the same net income, but that requires less time (e.g., less travel distance).
- A driver would always accept to do an extra distance *d* if he would get an extra net earnings of *d · (fare – vcost)*. This is actually the current rate a driver earns when accomplishing a service and, thus, he would always be willing to provide his service for this rate.

Let's consider that a driver $v$ is currently assigned to a service $s_k$ and the mediator wants the driver to do other services $s_j$ instead of $s_k$. Furthermore, let $td(v,s) = d(s_k) + d(v,s_k)$, denote the total distance driver $v$ has to accomplish in order to serve service $s$. In order to convince the driver, we define a compensation $c$ that is applied if a driver accepts the re-assignment. This compensation is calculated as follows:

Case1:  If $td(v,s_k) < td(v,s_j)$:  $c = [Earn(v, s_k) + (td(v,s_j) - td(v,s_k)) \cdot (fare - vcost)] - Earn(v, s_j)$

In this case the effective income of the driver when accepting the re-assignments and receiving the compensation $c$, would be $Earn(v, s_k) + (td(v,s_j) - td(v,s_k)) \cdot (fare - vcost)$. That is, the driver receives the same income as before, plus the normal fare for the extra distance.

Case2:  If $td(v,s_k) \geq td(v,s_j)$:  $c = Earn(v, s_k) - Earn(v, s_j)$

In this case the effective new income of the driver with the compensation is $Earn(v, s_k)$. Thus, the driver would have the same earnings as before, but for less distance (and less time).

It is clear that, with the assumptions mentioned above, an economically rational driver would accept any re-assignment with the defined compensations.

It should be noted that compensations may be positive or negative, e.g., a driver may get extra money for accepting a new service or s/he may have to pay some amount to the mediator. For instance, if a driver is proposed a re-assignment from a service $s_k$ to another service $s_j$ with $d(s_k) = d(s_j)$ and $d(v,s_k) > d(v,s_j)$ (case 2), the situation is a priori positive for the driver. S/he would have less distance to the starting point of the transportation service request but would earn the same money for the service itself. Thus, his net income would be higher. In this case, the compensation would be negative, with the amount $c = vcost \cdot (d(v,s_j) - d(v,s_k))$. That is, the driver would have to pay the cost of the difference in distances towards $s_j$ wrt. the previous service $s_k$.

The idea of the mediator is to dynamically find global re-assignments with compensations, such that the overall outcome of the mediator is cero or positive, e.g., there would be no extra mediation cost. Given an existing assignment $A_c$ at a given time, the algorithm we propose for calculating a new assignment $A_n$ with compensations is summarized as follows:

1. Assign all pending transportation requests to vehicles using the NVNS rule and add the assignments to $A_c$
2. Calculate an optimal assignment $A_n$ between all vehicles and requests assigned in $A_c$
3. Calculate the overall compensation $C_o$ to be paid/received to/from drivers for the change from $A_c$ to $A_n$
4. If *mediatorEarning - $C_o$ > 0* then
5.     *mediatorEarning := mediatorEarning - $C_o$*
6.     return $A_n$

7. else return $A_c$

The algorithm is executed by the mediator whenever either a new transportation request (service) is registered, or a driver becomes available (either after terminating a previous mission or because he starts working). In the first step, the system tries to assign pending requests in a rather standard fashion. Then, in step 2 and 3, a more efficient global assignment is searched for and the compensation cost of this new assignment is estimated. The new assignment is applied, if the accumulated overall mediator earnings together with the compensation cost remains positive. This last part assures that the mediator has no extra mediation cost.

Regarding step 2, we use Bertsekas' auction algorithm [47] to calculate an optimal assignment. In particular, we calculate the assignment $A_n$ that minimizes $D(A_n) + \gamma \cdot C_o$, where $D(A_n)$ is the sum of the distances of all vehicles in $A_n$ to the corresponding original positions of the assigned service requests. This means, we look for assignments that minimize the sum of the distances and also the potential cost of the compensations. $\gamma$ is a factor for scaling monetary earnings into distance values (meters).

### 3.5.2. Evaluation

We tested the proposed approach in different experiments simulating the operation of a taxi fleet which basically has the characteristics of the type of fleets we want to address here. We used an operation area of about 9×9 km, an area that roughly corresponds to the city centre of Madrid, Spain. In the simulations we randomly generated service requests (customers) who are assigned to available taxis, and we simulate the movement of taxis to pick up a customer, to drive him to his/her destination and then waiting for the assignment of a new customer. The simulations are not aimed at reproducing all relevant aspects of the real-world operation of a taxi fleet, but to analyse and compare the proposed coordination strategy (here called DYNRA) to the standard strategies FCFS and NVNR. Thus, we simplified the movements of taxis to straight-line movements with a constant velocity of 17 km/h. This velocity is within the range of the average velocity in the city centre of Madrid. Hence, we do not take into account neither the real road network, nor the possibility of different traffic conditions.

The general parameters used in the simulation are as follows. We use 1000 taxis (initially distributed randomly in the area with a uniform distribution) and a simulation interval of 5 hours. The taxis do not cruise, that is they only move if they are assigned to a customer. We accomplish different simulation runs with different numbers of customers in order to represent different supply/demand ratios. We generate a fixed number of customers every 15 minutes (ranging from 250 to 1000 in steps of 125). For each customer, his/her origin (point of appearance) and destination location are randomly chosen such that each trip goes either from the outside of the area to the centre, or vice versa. The origin and destination points are generated using a normal distribution (for centre and outside points). When a taxi arrives at a customer's location, a pick-up time of 30 seconds is used where the taxi does not move. In the same sense, the simulated drop-off time is 90 seconds. The system assignment process is accomplished every 5 seconds and only if a new client appeared or a taxi has become available again after a previous trip.

The payment scheme we used in the experiments is the one that has been used in the city of Madrid in the last years. A taxi trip has a fixed cost *fcost* = 2.4 euros and *fare* = 1.05 euros/km. Furthermore, the cost factor is *vcost* = 0.2 euros/km. This factor roughly corresponds to the actual cost of a vehicle, including petrol, maintenance, as well as other fixed costs. Finally, we apply a factor of $\gamma$=1/0.00085, which corresponds to the net benefit a taxi receives per meter when transporting a client in the used payment scheme.

Each experiment is repeated 10 times with a different random seed, in order to avoid biased results due to a particular distribution of clients. The presented results are averages over those 10 runs.

Table 2 presents the average waiting times of the customers for the three methods and the different numbers of generated customers per hour. As it can be observed, between 2000 and 2500 customers per hour, the FCFS approach starts to be perform really badly. Basically, the system gets saturated and the rate of serving customers is lower than the rate of appearance of new customers. The other two methods, NVNR and DYNRA, can deal much better with this situation and their saturation point is higher (between 2500 and 3000 customers per hour). There is a clear improvement in the waiting times of these two methods with respect to FCFS if there are more than 2000 customers per hour. The dynamic re-assignment approach with compensations performs better than the other two methods in all cases. The improvement is rather low if there are less customers but increases with the number of customers up to 101.4 and 2.4 minutes with respect to FCFS and NVNR, respectively, for 4000 customers a hour. In terms of relative improvement, the highest peak is reached at 2500 customers, with an improvement of 94.6% wrt. to FCFS and 44.9 % wrt. NVNR.

**Table 2.** Average waiting times for customers (in minutes).

|        | # customers per hour | | | | | | |
|--------|------|------|------|-------|-------|--------|--------|
| **Method** | **1000** | **1500** | **2000** | **2500** | **3000** | **3500** | **4000** |
| **FCFS**  | 1.2  | 1.56 | 1.92 | 39.02 | 75.29 | 112.01 | 148.01 |
| **NVNR**  | 1.2  | 1.56 | 1.92 | 3.83  | 8.29  | 27.38  | 49.01  |
| **DYNRA** | 1.19 | 1.48 | 1.73 | 2.11  | 6.83  | 25.24  | 46.62  |

It should be noted that the DYNRA approach is based on the compensation scheme presented above, that is, reassignments include compensations and (economically rational) taxi drivers will accept such re-assignments. In Figure 14 we analyse the net income of the system, composted of the income of the taxi drivers plus the income of the mediator in case of the DYNRA approach. The presented results are normalized to the income of 1000 drivers and 1000 customers. The overall system income is highest for the DYNRA approach for all numbers of customers. The difference to the FCFS approach is considerable, between 473 and 565 euros, above 2500 customers per hour (where FCFS is saturated). The difference of DYNRA wrt. NTNR is highest at 2500 customers (79 euros) and about 6-7 euros above that point. The taxi drivers earn always more money with the DYNRA approach up to 2500 customers per hour. However, their net income is slightly lower than in the NTNR approach for more customers. Nevertheless, it should be noted that the mediator could redistribute its income among all drivers and, thus, the drivers would have a higher income in all cases.

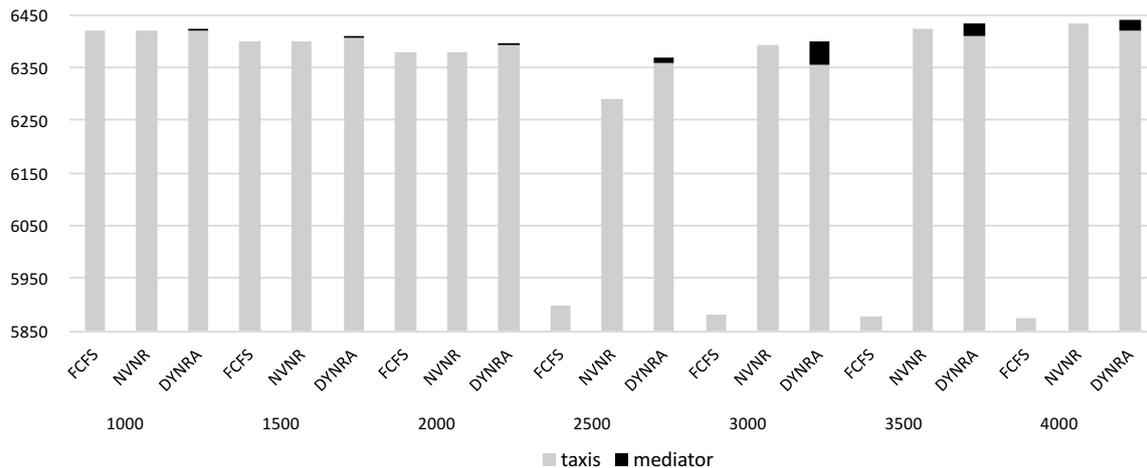

**Figure 14.** Average net income of taxi drivers and mediator in euros. The data are normalized to 1000 taxis serving 1000 customers.

Summarizing, the proposed dynamic re-assignment strategy can improve the performance of a transportation fleet of autonomous, self-interested drivers in terms of higher income and less movements (and thus, also more environmental friendly). The improvements are in general rather small if there are few movements (few service request) and higher if the demand of transportation services is increasing.

The approach relies on a mediator service that manages the assignments of transportation tasks to drivers and pays compensations if necessary. Besides this fact, as proposed here, the mediator does not incur in extra costs. Instead, it may have some positive income itself. The overall travel times and distances may still be reduced if the compensation system allowed for a negative balance. This could be of interest if, for example, a municipality would be willing to invest money in order to reduce $CO_2$ emissions.

## 4. Discussion

In this paper we have argued that recent technological advances open up new possibilities for computers to support people's interactions in a variety of domains with high socio-economic potential. In these domains, the choices and actions of a large number autonomous stakeholders need to be coordinated, and interactions can be regulated, by some sort of intelligent computing infrastructure, through institutions and institutional agents, or simply by providing information in an environment with a significant level of uncertainty. Many problems related to the vision of Smart Cities fall under this umbrella.

While centrally designed systems may be a suitable choice to address certain challenges related to Smart Cities, others are unlikely to be dealt with satisfactorily, either because stakeholders are *unwilling* to implement system recommendations that they do not understand and that they may not trust, or because it is *impossible* to compute good global solutions based on the information provided by stakeholders, which can be insufficient or biased by their personal interests. While the former problem can be addressed by providing stakeholders with their own trusted software agent which represents them and acts on their behalf, the latter requires coordination mechanisms that take into account the *autonomy* of the stakeholders (and their software agents). Still, designing and implementing such coordination mechanisms in open systems is challenging, especially if the systems are large in scale, as in the case of most smart Cities applications.

We argued that technologies from the AT sandbox are suitable for fostering coordination in such scenarios. To back this claim, we reported on a variety of real-world applications, ranging from truly open systems, where coordination among agents is achieved either though (economic) incentives or by offering relevant information, to more closed domains where AT techniques are used to "simulate" interactions among autonomous agents, which may, for instance, take the shape of auctions or market equilibria. While in some of the applications the use of AT, and market-based approaches in particular, was a mere design choice, in other more open domains their use enables the provision of new functionalities and services.

In fact, a key lesson learnt from our work is that market-based coordination schemes can be successfully applied to quite different problems and domains, even though their particular shape needs to be carefully tailored and adapted based on the degree of autonomy of the stakeholders. In the applications outlined in sections 3.3 and 3.4 the degree of autonomy of the different agents is low, as ambulance drivers, for instance, have to follow the assignments that they are given by the coordination mechanism. Evacuees in section 3.2 do have a choice but, due to the specific characteristics of the emergency situation and the scarceness of adequate information, the suggestions of the system are likely to be followed. A similar "take it or leave it" situation is present in the taxi fleet coordination example of section 3.5, but the stakeholders can make more informed decisions, so it is important that the incentives offered as part of the system proposal are such that taxi drivers can conclude that they are better off following the recommendations than without it. Finally, within the case study related to networks of reservation-based intersections in section 3.1 there are no explicit proposals for drivers to follow a particular route, but traffic assignment is achieved implicitly by coordinating the intersections' reserve prices. In addition, the auction protocol used at each reservation-based intersection needs to be such that the mechanism is resilient to attempts of strategic manipulation.

A limitation of our approach is related to the level of scalability needed for a particular domain. The applications outlined in this article have been evaluated in simulations with hundreds or thousands of agents, but going beyond these numbers may require the use of different approximations algorithms, e.g. to determine winner determination in auctions. Also, it should be noticed that take-it-or-leave-it recommendations may not work well when users can try to go for "outside options", i.e. when competing service providers exist and users are allowed to use them. Finally, making mechanisms stable against strategic manipulation attempts often relies on assumption regarding the underlying communication infrastructure, which may hold in simulation experiments but are not achievable in all real-world situations.

We intend to continue developing applications for the aforementioned type of domains making use of the AT sandbox. We will particularly be looking into applications for which the semantic as well as the trust and reputation layers of the AT tower are of foremost importance. This will help us broaden the set of models and tools based on AT. In much the same way, we plan to extract further guidelines for designing these type of systems, based on a descriptions of problems characteristics and requirements.

**Acknowledgments:** This work has been partially supported by the Spanish Ministry of Economy and Competitiveness through grant TIN2015-65515-C4-X-R ("SURF"), by the Autonomous Region of Madrid through grant S2013/ICE-3019 ("MOSI-AGIL-CM", co-funded by EU Structural Funds FSE and FEDER), and by URJC-Santander grant 30VCPIGI15. The authors would like to thank Dr. Matteo Vasirani for his contribution to the application for coordinating of traffic flows through intelligent intersections outlined in Section 3.1.

**Author Contributions:**

Holger Billhardt received his PhD in Informatics from the Technical University of Madrid in 2003. Since 2001, he works at the University Rey Juan Carlos in Madrid in the Artificial Intelligence Group, where he is currently an Associate Professor of Computer Science. His recent research has been concerned with multiagent systems, in particular with mechanisms for regulating the behaviour of software agents in open and dynamic environments. Prof. Billhardt is author or coauthor of more than 80 research papers and has participated in 19 nationally or internationally funded research projects and contracts. He has also been member of the organizing and programme committee of several international workshops and conferences.

Alberto Fernández is an Associate Professor of Computer Science and member of the Centre for Intelligent Information Technologies (CETINIA) at the University Rey Juan Carlos (URJC) in Madrid. He obtained his MSc degree in Informatics from the Technical University of Madrid (UPM) in 1999, and a PhD degree in Informatics from URJC in 2007. During the last years, he has worked on the fields of coordination in multiagent systems, semantic Web, and self-adaptive systems. He has published more than 100 research papers in international journals and conferences, and has participated in more than 20 research projects. He has been member of the organizing committee and the program committee of multiple international workshops and conferences.

Marin Lujak is an Associate Professor at IMT Lille Douai in France. He graduated Mechanical Engineering at the University of Zagreb, Croatia, in 2006 and received a PhD degree in Management Engineering in 2010 from the University of Rome "Tor Vergata", Italy. From 2011 to 2016, he was a Research Associate at the AI group of the University Rey Juan Carlos in Madrid, Spain. He has published more than 50 papers in internationally renowned journals, conferences and workshops. Moreover, he has participated as a referee in more than a dozen of renowned international journals and as a PC member and reviewer in more than 60 international conferences.

Sascha Ossowski is a Full Professor of Computer Science and the Director of the Centre for Intelligent Information Technologies (CETINIA) at University Rey Juan Carlos in Madrid. He obtained his MSc degree in Informatics from the University of Oldenburg (Germany) in 1993, and a PhD degree in Artificial Intelligence from UPM in 1997. He has authored more than 200 research papers, focusing on the application of Artificial Intelligence techniques to real world problems in various domains. He is co-editor of more than 20 books, proceedings, and special issues of international journals.

**References**

1. Ossowski, S. et al. *Agreement Technologies*. Law, Governance and Technology series (LGTS) no. 8. Springer, 2013, ISBN978-94-007-5582-6
2. Gelernter, D.; Carriero, N. Coordination Languages and their Significance. *Communications of the ACM* **1992**, Vol. 35, pp. 96-107
3. Ossowski, S.; Menezes, R. On coordination and its significance to distributed and multi-agent systems. *Concurrency and Computation: Practice and Experience* **2006**, Vol. 18, pp. 359-370
4. Vito, A.; Berardi, U.; Dangelico, R. Smart cities: Definitions, dimensions, performance, and initiatives. *Journal of Urban Technology* **2015**, 22.1, pp. 3-21.
5. Wooldridge, M., Jennings, N. Intelligent Agents -- Theory and Practice. *Knowledge Engineering Review* **1995**, Vol. 10, pp. 115-152
6. Sierra, C., Botti, V., Ossowski, S: Agreement Computing. *Künstliche Intelligenz* **2011**, Vol. 25, pp. 57--6
7. Ossowski, S.; Sierra, C.; Botti, V. Agreement Technologies --A Computing perspective. In: [1]. pp 5-18
8. Schumacher, M.; Ossowski, S. The Governing Environment. In: *Environments for Multi-Agent Systems II (E4MAS)*. Springer, 2005, pp. 88-104
9. Omicini, A.; Ossowski, S.; Ricci, A. Coordination Infrastructures in the Engineering of Multiagent Systems. In: *Methodologies and software engineering for agent systems – The Agent-Oriented Software Engineering Handbook* (Bergenti, Gleizes, y Zambonelli, editores). Springer, 2004, pp. 273-296, ISBN: 1-4020-8057-3.
10. Ossowski, S.: Coordination in Multi-Agent Systems -- Towards a Technology of Agreement. In: *Multiagent System Technologies (MATES-2008)*. LNCS 5244, Springer, 2008, pp. 2-12
11. Ossowski, S.: Constraint Based Coordination of Autonomous Agents. *Electronic Notes Theoretical Computer Science* **2001**, Vol. 48, pp. 211-226
12. Von Martial, F.: *Co-ordinating Plans of Autonomous Agents*. LNAI 610, Springer, 1992
13. Shoham, Y., Leyton-Brown, L.: *Multiagent Systems – Algorithmic, Game-Theoretic, and Logical Foundations*. Cambridge Univ. Press, 2009
14. Fornara, N.; Lopes Cardoso, H.; Noriega, P.; Oliveira, E.; Tampitsikas, C. Modelling Agent Institutions. In [1], pp. 277-308
15. Centeno, R.; Billhardt, H.; Hermoso, R.; Ossowski, S. Organising MAS – A formal model based on organisational mechanisms. In: *Proceedings of the 2009 ACM symposium on Applied Computing (SAC)*, 2009, pp. 740-746
16. Nam, T.; Pardo, T. A. Conceptualizing smart city with dimensions of technology, people, and institutions. In: *Proc. 12th annual international digital government research conference: digital government innovation in challenging times*, 2011, ACM, pp. 282-291.
17. Giffinger, R.; Fertner, C.; Kramar, H.; Kalasek, R.; Pichler-Milanovic, N.; Meijers, E. (2007). Smart cities- Ranking of European medium-sized cities. Vienna University of Technology. http://www.smartcities.eu/download/smart_cities_final_report.pdf.
18. O'Grady, M.; O'Hare, G. How smart is your city? *Science* 2012, Vol. 335, pp. 1581-1582.

19. Ahvenniemi, H.; Huovila, A.; Ointo-Seppä, I.; Airaksinen, M. What are the differences between sustainable and smart cities? *Cities (Part A)* **2016**, Vol 60, pp. 234-245.
20. Bibri, S.; Krogstie, J. Smart sustainable cities of the future: An extensive interdisciplinary literature review." *Sustainable Cities and Societies* **2017**, Vol. 31, pp. 183-212.
21. Kondepudi, S. N.; Ramanarayanan, V.; Jain, A.; Singh, G. N.; Nitin Agarwal; N. K., Kumar, R.; ... ; Gemma, P. (2014). *Smart sustainable cities: an analysis of definitions*. The ITU-T Focus Group for Smart Sustainable Cities. https://www.itu.int/en/ITU-T/focusgroups/ssc
22. Mohanty, S.; Choppali, U.; Kougianos, E.. Everything you wanted to know about smart cities: The internet of things is the backbone. *IEEE Consumer Electronics Magazine* **2016**, Vol. 5, pp. 60-70.
23. Petrolo, R.; Loscri, V.; Mitton, N.. Towards a smart city based on cloud of things, a survey on the smart city vision and paradigms. *Transactions on Emerging Telecommunications Technologies* **2015**, Vol. 28
24. Dresner, K.; Stone, P. A Multiagent Approach to Autonomous Intersection Management. *Journal of Artificial Intelligence Research* **2008**, Vol. 31, pp. 591-656
25. Vasirani, M.; Ossowski, S. Evaluating Policies for Reservation-Based Intersection Control. In: *Proceedings of the 14th Portuguese Conference on Artificial Intelligence*, 2009, pp. 39-50
26. Schepperle, H.; Bohm, K. Agent-based traffic control using auctions. In: *Cooperative Information Agents XI (CIA-2007)*, LNCS 4676, Springer, pp. 119–133
27. Vasirani, M.; Ossowski, S. A market-inspired approach for intersection management in urban road traffic networks. *Journal of Artificial Intelligence Research* **2012**, Vol. 43, pp. 621–659
28. Vasirani, M.; Ossowski, S. Learning and Coordination for Autonomous Intersection Control. *Applied Artificial Intelligence* **2011**, Vol. 25, pp 193-216
29. Codenotti, B.; Pemmaraju, S.; Varadarajan, K. The computation of market equilibria. *SIGACT News* **2004**, Vol. 35, pp. 23-37
30. Avery, W.H.; Soo,J. Emergency/disaster guidelines and procedures for employees. CCH Canadian Limited, 2003
31. Chen, P.H., Feng, F.: A fast flow control algorithm for real-time emergency evacuation in large indoor areas. Fire Safety Journal **2009**, Vol. 44(5), pp. 732–740
32. Filippoupolitis, A., Gelenbe, E.: A distributed decision support system for building evacuation. In: 2nd IEEE Conference on Human System Interactions, 2009, pp. 323–330
33. Lujak, M., Ossowski, S.: Intelligent people flow coordination in smart spaces. In: *Multi-Agent Systems and Agreement Technologies* (EUMAS'15 and AT'15), Springer, 2016, pp. 34-49
34. Etzion, O.; Niblett, P. *Event Processing in Action*, Manning Publications, 2010. ISBN 1935182218.
35. Khaleghi, B., Khamis, A., Karray, F.O., Razavi, S.N. Multi-sensor data fusion – A review of the state-of-the-art. *Information Fusion* **2013**, Vol. 14, pp. 28–44
36. Zervas, E., Mpimpoudis, A., Anagnostopoulos, C., Sekkas, O., Hadjiefthymiades, S.: Multi-sensor data fusion for fire detection. *Information Fusion* **2011**, Vol. 12, pp. 150–159
37. Lujak, M., Giordani, S., Ossowski, S.: Route guidance: Bridging system and user optimization in traffic assignment. *Neurocomputing* **2015**, Vol. 151, pp. 449–460
38. Lujak, M.; Giordani, S. Centrality measures for evacuation: Finding agile evacuation routes. *Future Generation Computer Systems* **2017**, Vol. 83, pp. 401-412
39. Rajagopalan, H. K.; Saydam, C.; Xiao, J. A multiperiod set covering location model for dynamic redeployment of ambulances, *Comput. Oper. Res.* **2008**, Vol. 35 (3), pp. 814–826
40. Maxwell, M. S.; Restrepo, M.; Henderson, S. G.; Topaloglu, H. Approximate dynamic programming for ambulance redeployment *INFORMS J. Comput.* **2010**, Vol. 22 (2), pp. 266–281
41. Naoum-Sawaya, J.; Elhedhli, S. A stochastic optimization model for real-time ambulance redeployment, *Comput. Oper. Res.* **2013**, Vol. 40 (8), pp. 1972–1978
42. Andersson, T.; Varbrand, P. Decision support tools for ambulance dispatch and relocation, *J. Oper. Res. Soc.* **2007**, Vol. 58 (2), pp. 195–201
43. Bandara, D.; Mayorga, M.E.; McLay, L.A. Priority dispatching strategies for EMS systems, *Journal of the Operational Research Society* **2013**, Vol. 65 (4), pp. 572–587
44. Billhardt, H.; Lujak, M.; Sánchez-Brunete, V., Fernández, A.; Ossowski. S. Dynamic coordination of ambulances for emergency medical assistance services. *Knowledge Based Systems* **2014**, Vol. 70., pp 268-280
45. Haghani, A; Hu, H.; Tian, Q. An optimization model for real-time emergency vehicle dispatching and routing. In: 82nd Annual Meeting of the Transportation Research Board, Washington, DC, 2003

46. Munkres, J. Algorithms for the assignment and transportation problems, *J. Soc. Indus. Appl. Math.* **1957**, Vol. 5, pp. 32–38
47. Bertsekas, D. The auction algorithm: a distributed relaxation method for the assignment problem. *Ann. Oper. Res.* **1988**, Vol. 14, pp. 105–123
48. Du, Q. ; Faber, V. ; Gunzburger, M. Centroidal voronoi tessellations: applications and algorithms, *SIAM Rev.* **1999**, Vol. 41, pp. 637–676
49. Lloyd, S. Least squares quantization in PCM. *IEEE Trans. Inform. Theory* **1982**, Vol. 28, pp. 129–137
50. *WHO Fact sheet No. 310*, World Health Organization, 2015
51. Van De Werf, F.; Bax, J.; Betriu, A. et al.. Management of acute myocardial infarction in patients presenting with persistent st-segment elevation, *European heart journal* **2008**, Vol. 29, pp. 2909-2945
52. Wilde, E.T. Do emergency medical system response times matter for health outcomes, *Health economics* **2013**, Vol. 22, pp. 790-806
53. Lujak, M.; Billhardt, H.; Ossowski, S. Distributed Coordination of Emergency Medical Assistance for Angioplasty Patients. *Annals of Mathematics and Artificial Intelligence* **2016**, Vol. 78, pp. 73-100
54. Lujak, M.; H. Billhardt, H. Coordinating Emergency Medical Assistance. In [1], pp. 597-509
55. Guerriero, F.; Guido, R. Operational research in the management of the operating theatre: A survey. *Health Care Manag. Sci.* **2011**, Vol. 14, pp. 89-114
56. Chevaleyre, Y.; Endriss, U.; Estivie, S.; Maudet, N. Multiagent resource allocation in k-additive domains: preference representation and complexity. *Annals OR* **2008**, Vol. 163, pp. 49-62
57. Endriss, U.; Maudet, N.; Sadri, F.; Toni, F. On optimal outcomes of negotiations over resources. In: *Proc. Joint Intl Conf. on Multiagent Systems (AAMAS)*, 2003, pp. 177-184.
58. Billhardt, H.; Fernández, A.; Lujak, M.; Ossowski, S.; Julián, V.; De Paz, J.F., Hernandez, J.Z. Coordinating open fleets. A taxi assignment example. AI Communications **2017**, Vol. 30, pp. 37-52
59. Maciejewski, M.; Bischoff, J.; Nagel, K. An Assignment-Based Approach to Efficient Real-Time City-Scale Taxi Dispatching, IEEE Intelligent Systems **2016**, Vol. 31, pp. 68–77
60. Zhu, C., & Prabhakar, B. Reducing Inefficiencies in Taxi Systems. In: *Proc. of the 56th IEEE Conference on Decision and Control (CDC)*, 2017
61. Dai, G., Huang, J., Wambura, S. M., & Sun, H. A Balanced Assignment Mechanism for Online Taxi Recommendation. In: *Proc. of the 18th IEEE International Conference on Mobile Data Management (MDM)*, 2017, pp. 102–111
62. Gao, G., Xiao, M., & Zhao, Z. Optimal Multi-taxi Dispatch for Mobile Taxi-Hailing Systems. In: *Proc. of the 45th International Conference on Parallel Processing (ICPP)*, 2016, pp. 294-303.
63. M. Meghjani, and K. Marczuk. A hybrid approach to matching taxis and customers. In: *Proc. of the Region 10 Conference (TENCON)*, 2016, pp. 167-169
64. Glaschenko, A., Ivaschenko, A., Rzevski, G., & Skobelev, P. Multi-Agent real time scheduling system for taxi companies. *In Proc. Int. Conf. Autonomous. Agents and Multiagent Systems (AAMAS)*, 2009, pp. 29-36